\documentclass[lettersize,journal]{IEEEtran}
\usepackage{amsmath,amsfonts,amssymb,amsthm}
\usepackage{algorithmic}
\usepackage{algorithm}
\usepackage{array}
\usepackage[caption=false,font=normalsize,labelfont=sf,textfont=sf]{subfig}
\usepackage{textcomp}
\usepackage{stfloats}
\usepackage{url}
\usepackage{verbatim}
\usepackage{graphicx}
\usepackage{threeparttable}
\newcolumntype{P}[1]{>{\centering\arraybackslash}p{#1}}
\usepackage{cite}
\usepackage{enumitem}
\usepackage[bottom]{footmisc}
\usepackage{balance}
\usepackage{hyperref}
\hypersetup{hidelinks=true}
\newtheorem{theorem}{Theorem}
\newtheorem{prop}{Proposition}

\usepackage{eso-pic}

\AddToShipoutPictureBG*{%
  \AtPageUpperLeft{%
    \put(40,-30){%
      \begin{minipage}{\textwidth}
        \footnotesize\centering
        This work has been submitted to the IEEE for possible publication. Copyright may be transferred without notice, after which this version may no longer be accessible.
      \end{minipage}%
    }%
  }%
}

\begin{document}

\title{Robust Adaptive Lyapunov-Based MPC for Nonlinear Systems with Experimental Validation on Quadcopter Attitude Control}

\author{Teo~Protoulis,
        Alex~Alexandridis\textsuperscript{*},~\IEEEmembership{Senior Member,~IEEE}
\thanks{Authors are with the University of West Attica, Department of Electrical and Electronic Engineering, Athens, 12241, Greece. \{tprotoulis, alexx\}@uniwa.gr}
\thanks{\textsuperscript{*}Corresponding author.}}

\markboth{Journal of \LaTeX\ Class Files,~Vol.~XX, No.~X, XXXX~XXXX}%
{Protoulis \MakeLowercase{\textit{et al.}}: Robust Adaptive Lyapunov-Based MPC for Nonlinear Systems with Experimental Validation on Quadcopter Attitude Control}


\maketitle

\begin{abstract}
In this paper, a computationally efficient robust adaptive Lyapunov-based model predictive control (RALMPC) framework is proposed for a class of constrained nonlinear systems subject to uncertainties and external disturbances. A radial basis function neural network (RBFNN), introduced to approximate non-parametric uncertainties, is incorporated into a control Lyapunov function analysis, together with integral barrier Lyapunov functionals for state constraint enforcement. This analysis yields control and adaptation laws for parametric uncertainties, external disturbances, RBFNN weights, and worst-case approximation error, rendering the closed-loop system uniformly bounded while ensuring convergence of tracking errors to compact sets around the origin. The resulting guarantees are then embedded into the RALMPC optimal control problem through a contractive constraint, thus combining formal closed-loop stability with real-time performance optimization; at the same time, the adaptation laws allow the RALMPC prediction model to be updated online. Comparative hardware-in-the-loop simulations against state-of-the-art MPC schemes validate the efficacy and computational efficiency of the proposed framework. Ultimately, RALMPC is experimentally validated on a custom-built quadcopter platform for time-varying attitude tracking, demonstrating its practical feasibility; comparisons against alternative strategies confirm its superiority in achieving improved closed-loop performance.

\end{abstract}


\section{Introduction}
\IEEEPARstart{I}{n} recent years, the control of nonlinear dynamical systems has increasingly been driven by the dual objective of guaranteeing safety-critical properties while simultaneously achieving high closed-loop performance. As autonomous systems become more complex and are deployed in uncertain, real-world environments, the performance requirements placed on control algorithms continue to intensify. Modern controllers are therefore expected not only to stabilize nonlinear dynamics, but also to enforce state and input constraints, ensure robustness against disturbances and model uncertainties, and maintain reliable operation under safety-critical conditions.

In this context, control barrier functions (CBFs) have emerged as a powerful state-of-the-art framework to enforce safety-critical constraints~\cite{ames2017,xwu2022}. By encoding safety requirements as forward invariance conditions, CBFs enable the systematic construction of control frameworks that guarantee constraint satisfaction by solving low-complexity quadratic programs. They have been successfully applied to a wide range of safety-critical applications, including robotic systems, autonomous vehicles, and multi-agent coordination~\cite{lwang2017,ames2019}. However, CBF-based designs typically act as supervisory filters around a nominal controller, modifying its control action to ensure safety. Therefore, while safety is rigorously enforced, performance of the nominal controller may be partially sacrificed. Moreover, undesired stable equilibria may be induced~\cite{mreis2021}, their avoidance generally requiring additional compatibility conditions~\cite{mreis2026}. Finally, the synthesis and composition of CBFs become increasingly intricate when the system is subject to both parametric and non-parametric uncertainties, and external disturbances.

An alternative and widely adopted framework for simultaneously addressing safety and performance objectives is model predictive control (MPC) \cite{rawlings2017}. Unlike supervisory safety filters, MPC incorporates constraints directly into a finite-horizon optimal control problem (OCP), allowing safety requirements and performance criteria to be treated within a unified optimization framework. By repeatedly solving this OCP in a receding-horizon manner, MPC can explicitly balance constraint satisfaction, robustness considerations, and closed-loop performance. This ability to integrate safety-critical constraints with performance optimization has made MPC a dominant methodology in nonlinear control applications.

Despite its strong theoretical foundation, MPC design becomes considerably more complex in the presence of uncertainty. Parametric variations, unmodeled dynamics, and external disturbances pose significant challenges in guaranteeing recursive feasibility and  constraint satisfaction. To address these issues, robust tube-based MPC (RTMPC) \cite{villanueva2017,sfeng2020,nezami2025} constructs a nominal optimal trajectory while an ancillary feedback law confines the true system trajectory within a robust invariant tube \cite{kohler2021,vicente2023,khuo2026}. By tightening the nominal constraints, RTMPC ensures feasibility of the actual state and input trajectories under the worst-case uncertainty realizations.

Although RTMPC offers strong theoretical guarantees, its practical implementation presents notable challenges. It requires explicit a priori bounds on uncertainties and disturbances to construct invariant tubes, which can be restrictive, if at all possible, particularly for nonlinear systems. Moreover, enforcing constraint satisfaction under worst-case realizations introduces conservatism, leading to excessive constraint tightening, potential feasibility loss, and increased prediction horizons to satisfy terminal constraints. The additional computations associated with tube construction and invariant set generation further inflate the computational burden, limiting the applicability of RTMPC to processes with relatively slow sampling rates \cite{khuo2026,xliu2023,clee2024,svensen2023,franze2023}.

To reduce conservatism and enhance performance, robust adaptive tube-based MPC (RATMPC) schemes have been proposed \cite{xlu2019,lorenzen2019,kzhang2020}. These approaches retain the guarantees of RTMPC while updating uncertainty sets online \cite{kohler2019a}, thereby alleviating excessive constraint tightening. However, this improvement comes at the cost of increased computational complexity: online adaptation enlarges the decision space, and additional terminal conditions are typically required to ensure recursive feasibility and stability. Therefore, deployment in fast dynamical systems becomes even more challenging.

Furthermore, RATMPC is well established for linear systems \cite{kohler2019a,kzhang2020,adey2025}, but its extension to nonlinear dynamics remains limited due to the complexity of constructing robust invariant tubes \cite{kohler2021a,sasfi2023}. Existing approaches rely on local linear prediction models \cite{lwu2022,strelnikova2025} and primarily address setpoint regulation rather than time-varying tracking, while real-time deployment is restricted to systems with slow sampling rates, frequently without explicit closed-loop guarantees \cite{casagrande2025,cliu2026}. A notable experimental validation on quadrotor altitude control is reported in \cite{didier2021}; however, it employs a simplified 2-state linear model, while the fast inner-loop attitude dynamics are handled by conventional PID controllers.
 
As an alternative approach for designing stabilizing MPC schemes, Lyapunov-based MPC (LMPC) has gained increasing attention \cite{mhaskar2006}. In contrast to conventional MPC formulations, LMPC enforces closed-loop stability through a contractive constraint imposed at the first step of the prediction horizon, constructed on the basis of a controller tailored to the particular setup under consideration \cite{christofides2013,ldai2023}. This formulation alleviates the need for explicitly designing terminal ingredients to guarantee stability and feasibility. In this respect, LMPC reduces computational complexity while combining the rigorous guarantees of Lyapunov-based control with MPC performance optimization, allowing improved closed-loop performance under formal specifications \cite{napolitano2022,zdu2024,bxu2025,protoulis2026scl,protoulis2026aerospace}. However, a critical issue in LMPC is to ensure compliance between the state and input constraints that are integrated into the OCP with the contractive constraint. Failing to do so results in the loss of recursive feasibility and stability guarantees.

Motivated by the aforementioned, in this paper, a robust adaptive LMPC (RALMPC) framework for time-varying output tracking in a class of uncertain nonlinear constrained systems is proposed; the main contributions are:\begin{itemize}[leftmargin=*]

    \item An integral barrier Lyapunov function (iBLF)-based adaptive control protocol is developed without requiring prior knowledge of system parameters or internal dynamics. All closed-loop signals are proven uniformly bounded, while the tracking error converges to a compact neighborhood of the origin under guaranteed state-constraint satisfaction. Parametric uncertainties and external disturbances are compensated via adaptation laws, and an adaptive RBFNN is employed to estimate unmodeled state-dependent dynamics.

    \item Based on the derived control and adaptation laws, a Lyapunov-based MPC framework is constructed that incorporates a contractive constraint to ensure recursive feasibility and inherit the closed-loop properties of the baseline controller. Unlike RATMPC and other alternative robust MPC schemes, this alleviates the need for designing robust invariant tubes and terminal ingredients, requiring a-priori knowledge of uncertainty bounds, or relying on Lipschitz constants to establish closed-loop formality. Furthermore, the adaptation laws are embedded within the MPC prediction model, enabling online model refinement. In addition, to account for actuation bounds, a relaxed polynomial barrier function (RPBF) is introduced that heavily penalizes controls outside the admissible region, while leaving a dead-zone to account for the unknown varying control input equilibrium.

    \item Compared to standard LMPC designs, our approach eliminates the need for a-priori characterization of the stability region \cite{mhaskar2006}; furthermore, uncertainty handling is directly embedded into the OCP via the contractive constraint and the adaptive prediction model without requiring external disturbance observers \cite{hwei2021}.

    \item The proposed RALMPC controller effectively reduces computational complexity, enabling real-time deployment on fast dynamical systems and resource-constrained embedded platforms. We demonstrate its practical benefits using hardware-in-the-loop simulations, followed by experiments on the attitude control of a physical quadcopter unit, including robust tracking under external disturbances.
    
\end{itemize}

The remainder of this manuscript is organized as follows. In section \ref{sec:problem-form}, the system dynamics and problem formulation are presented, while section \ref{sec:lyapunov-control} introduces the Lyapunov-based design. The developed RALMPC framework is presented in section \ref{sec:ralmpc-design}, the simulation and experimental results are included in section \ref{sec:exp-sim-scenarios}, and finally concluding remarks together with future research directions are stated in section \ref{sec:conclusions}.

\section{Problem Formulation}
\label{sec:problem-form}

\subsection{System Dynamics}
\label{subsec:sys-dynamcis}
We consider single-input single-output systems of the form:\begin{align}
    \dot{x}_1 &= x_2  \label{eq:system-dynamics} \\ 
    \dot{x}_2 &= f_n(x)+b\cdot u+F(x) + d(t), &
    y &=x_1 \notag
\end{align}where $x=[x_1 \ x_2]^T$ is the state vector, $u\in\mathbb{R}$ is the control input, $b$ is a control gain that is considered unknown, $f_x(x)$ is a smooth, locally Lipschitz function representing the nominal dynamics, $F(x)$ is a smooth, locally Lipschitz function denoting the unknown dynamics, $d(t)$ corresponds to the unknown disturbance signal, and $y$ is the system output.

\textit{Assumption 1}: The state $x$ is available for measurement.

\textit{Assumption 2}: The control gain $b$ is sign-definite and its sign is known. In the subsequent analysis, we assume $b>0$.

\textit{Assumption 3}: The disturbance signal $d(t)$ is piecewise constant (or slowly time-varying) and bounded, i.e., $\dot{d}(t)\approx 0, |d(t)|\leq \bar{D}$ for some unknown constant $\bar{D}>0$.

\textit{Remark 1}: The analysis focuses on systems of the form \eqref{eq:system-dynamics}, but the framework readily extends to higher-order systems in controller canonical form. A nominal model $f_n(x)$ is assumed when available; otherwise, the same design applies, by absorbing all state-dependent dynamics into $F(x)$.


\subsection{Preliminaries}
\label{subsec:preliminaries}
To approximate the unknown dynamics $F(x)$ of \eqref{eq:system-dynamics}, we employ an RBFNN with fixed centers and widths, employing a Gaussian activation function \cite{patrinos2010}. Given an RBFNN with $L$ nodes, there exists an ideal set of synaptic weights $w^*\in\mathbb{R}^{L\times1}$ that can approximate $F(x)$ within a compact domain $\mathcal{D}\subset\mathbb{R}^2$ to a certain degree of accuracy:\begin{equation}
    F(x)=Z(x)^T\cdot w^{*} + \epsilon(x) \label{eq:rbfnn-approximation}
\end{equation}where $Z(x)\in\mathbb{R}^{L\times 1}$ is the RBFNN hidden layer output and $\epsilon(x)$ is the approximation error that satisfies $|\epsilon(x)|\leq\bar{\epsilon}, \bar{\epsilon}>0$. Define the parametric error $\tilde{w}=\hat{w}-w^*$, where $\hat{w}$ denotes the weights' estimates, and substitute into \eqref{eq:rbfnn-approximation};\begin{equation}
    F(x)=Z(x)^T\cdot (\hat{w}-\tilde{w})+\epsilon(x) \label{eq:rbfnn-approx-with-tilde-w}
\end{equation}Define the parametric error $\tilde{\epsilon}=\hat{\epsilon}-\bar{\epsilon}$ with $\hat{\epsilon}$ being the estimate of $\bar{\epsilon}$ and $\tilde{b}=\hat{b}-b$, where $\hat{b}$ denotes the estimate of the control gain. Under the aforementioned, $\dot{x}_2$ can be written as:\begin{equation}
    \dot{x}_2=f_n(x)+\hat{b}u+Z(x)^T\hat{w}-\tilde{b}u-Z(x)^T\tilde{w}+\epsilon(x) + d(t) \label{eq:augmented-x2dot}
\end{equation}

\subsection{Control Problem Formulation}
\label{subsec:control-prob-form}
Consider \eqref{eq:system-dynamics} and define the output tracking error $z_1(t)=x_1(t){-}r(t)$, where $r(t)$ denotes a time-varying reference. Design a feedback control policy that ensures:\begin{itemize}[leftmargin=*]
    \item Convergence of the tracking error to a compact set around the origin, whose size can be reduced through selection of the controller parameters.
    \item Strict satisfaction of state constraints of the form $|x_1(t)|<\bar{x}_1,|x_2(t)|<\bar{x}_2,\forall t\geq 0$.
    \item System controllability is maintained, i.e., $\hat{b}(t)>0,\forall t\geq 0$.
    \item The resulting optimal control framework accounts for explicit control input bounds without compromising feasibility.
    \item All closed-loop signals remain bounded.
\end{itemize}

The presence of state constraints has a twofold objective: i) ensure safety-critical operation of the controlled system, and ii) maintain the closed-loop operation within the compact domain $\mathcal{D}:=[-\bar{x}_1 \ \bar{x}_1]\times [-\bar{x}_2 \ \bar{x}_2]\subset \mathbb{R}^2$, where the RBFNN approximation is valid. The latter is critical as in case the state leaves $\mathcal{D}$, then the RBFNN is asked to extrapolate and the estimates of $F(x)$ are no longer reliable, potentially leading to control failure.

\textit{Assumption 4}: The reference trajectory is at least $\mathcal{C}^2$ with $|r(t)|,|\dot{r}(t)|,|\ddot{r}(t)|\leq\bar{r},\dot{\bar{r}},\ddot{\bar{r}}$. Moreover, $r(t),\dot{r}(t)$ are strictly feasible w.r.t. to the imposed state constraints.

\section{Lyapunov Design}
\label{sec:lyapunov-control}
\subsection{Control and Adaptation Laws}
\label{subsec:lyap-control-and-adaptation-laws}
To derive the control and adaptation laws, we follow a backstepping procedure similar to \cite{tang2016}. However, we do not require specific knowledge of the bounds for $b$, but rather estimate it and later use a projection operator only to ensure the ultimate boundedness of $\hat{b}$. The latter is important, as we simultaneously ensure formal closed-loop properties and derive an adaptation law for $b$, which, together with the RBFNN and the disturbance estimation laws, are used to update the RALMPC prediction model online (see section \ref{sec:ralmpc-design}).

\textit{$1^{\text{st}}$ Step}: To enforce $|x_1(t)|<\bar{x}_1$, define the iBLF\begin{equation}
    V_{x_1}(z_1,r):=\int\limits_{0}^{z_{1}}\frac{\sigma\cdot\bar{x}_1^2}{\bar{x}_1^2-(\sigma+r)^2}d\sigma \label{eq:x1-constraint-functional}
\end{equation}which is positive definite w.r.t. $e_1$ and satisfies $ V_{x_1}{\to}\infty \text{ as } |x_1|{\to}\bar{x}_1 $; its time derivative is:\begin{equation*}
    \dot{V}_{x_1}=\frac{\partial V_{x_1}}{\partial z_1}\dot{z}_1+\frac{\partial V_{x_1}}{\partial r}\dot{r} = \frac{\bar{x}_1^2z_1}{\bar{x}_1^2-x_1^2}\cdot x_2 - e_1\rho_1(z_1,r)\dot{r}
\end{equation*}where $\rho_1(z_1,r)=\int_{0}^{1}\frac{\bar{x}_1^2}{\bar{x}_1^2-(\theta z_1+r)^2}d\theta$. Introduce the auxiliary error metric $z_2=x_2-a_1$, where $a_1$ is a virtual control function to be designed and $\dot{V}_{x_1}$ becomes\begin{equation}
    \dot{V}_{x_1} = \frac{\bar{x}_1^2z_1}{\bar{x}_1^2-x_1^2}\cdot (z_2+a_1) - z_1\rho_1(z_1,r)\dot{r} \label{eq:first-step-firs-lyap-der}
\end{equation}Select the virtual control function as\begin{equation}
    a_1=-k_1z_1+\frac{\bar{x}_1^2-x_1^2}{\bar{x}_1^2}\rho_1(z_1,r)\dot{r}, \ k_1>0 \label{eq:first-step-virtual-control}
\end{equation}Substituting \eqref{eq:first-step-virtual-control} into \eqref{eq:first-step-firs-lyap-der} yields\begin{equation}
    \dot{V}_{x_1}=-\overbrace{k_1\frac{\bar{x}_1^2z_1^2}{\bar{x}_1^2-x_1^2}}^{W(x_1,r)\geq0} + \frac{\bar{x}_1^2z_1z_2}{\bar{x}_1^2-x_1^2} \label{eq:first-step-final-lyap-der}
\end{equation}which contains a negative definite term in \(z_1\) and a coupling term involving \(z_2\), to be handled in the next backstepping step.

\textit{$2^{\text{nd}}$ Step}: Define the iBLF for constraint $|x_2(t)|<\bar{x}_2$:\begin{equation}
    V_{x_2}(z_2,a_1)=\int\limits_{0}^{z_2}\frac{\sigma\bar{x}_2^2}{\bar{x}_2^2-(\sigma+a_1)^2}d\sigma \label{eq:second-step-iblf}
\end{equation}To improve readability, let us define the following terms:\begin{align}
    &\rho_2(z_2,a_1)=\int\limits_{0}^{1}\frac{\bar{x}_2^2}{\bar{x}_2^2-(\theta z_2+a_1)^2}d\theta \label{eq:rho2-definition} \\
    &s_1 =\frac{\bar{x}_1^2z_1}{\bar{x}_1^2-x_1^2}, \quad
    s_2=\frac{\bar{x}_2^2z_2}{\bar{x}_2^2-x_2^2} \label{eq:log-barrier-scaled-term}
\end{align}Then, the time derivative of $V_{x_2}$ is\begin{equation}
    \dot{V}_{x_2} = s_2\dot{x}_2-z_2\rho_2(z_2,a_1)\dot{a}_1 \label{eq:x2-iblf-derivative}
\end{equation}Based on the above definition, the parametric errors defined in section \ref{subsec:preliminaries} and $\tilde{d}(t)=\hat{d}(t)-d(t)$, the composite candidate Lyapunov function is formulated as\begin{equation}
    V_c(x,r)=V_{x_1}+V_{x_2}+\frac{\tilde{b}^2}{2\gamma_1}+\frac{\tilde{w}^T\tilde{w}}{2\gamma_2}+\frac{\tilde{\epsilon}^2}{2\gamma_3} +\frac{\tilde{d}^2}{2\gamma_4}\label{eq:composite-lyapunov-function}
\end{equation}where $\gamma_i>0,i=1,\ldots,4$ are adaptation rates. The time derivative of $V_c$ is by virtue of \eqref{eq:augmented-x2dot}, \eqref{eq:first-step-final-lyap-der} and \eqref{eq:x2-iblf-derivative}:\begin{align}
    &\dot{V}_c = -W(x_1,r)+s_1z_2+s_2\big[f_n(x)+\hat{b}u+Z(x)^T\hat{w}+\hat{d}\big] \notag \\
    &-z_2\rho_2\dot{a}_1 + \tilde{b}\big(\gamma_1^{-1}\dot{\hat{b}}-s_2u\big) + \tilde{w}^T\big(\gamma_2^{-1}\dot{\hat{w}} - s_2Z(x)\big) \notag \\ &+ \tilde{d}(\gamma_4^{-1}\dot{\hat{d}}-s_2) 
    +\gamma_3^{-1}\tilde{\epsilon}\dot{\hat{\epsilon}} + s_2\epsilon(x)= \label{eq:second-step-lyap-der} \end{align}\begin{align}
    &=-W(x_1,r)+z_2\Big[s_1+\frac{\bar{x}_2^2}{\bar{x}_2^2-x_2^2}\big[f_n(x) + \hat{b}u+Z(x)^T\hat{w}\notag \\
    & + \hat{d}\big] -\rho_2\dot{a}_1\Big] + \tilde{b}\big(\gamma_1^{-1}\dot{\hat{b}}-s_2u\big) + \tilde{w}^T\big(\gamma_2^{-1}\dot{\hat{w}} - s_2Z(x)\big) \notag \\ &+ \tilde{d}(\gamma_4^{-1}\dot{\hat{d}}-s_2)+\gamma_3^{-1}\tilde{\epsilon}\dot{\hat{\epsilon}} + s_2\epsilon(x)\notag
\end{align}To proceed we select the control input as\begin{equation}
    u(t)=\hat{b}(t)^{-1}\cdot\big[ u_n(t)+u_r(t) \big] \label{eq:control-summation}
\end{equation}where $u_n(t)\in\mathbb{R}$ denotes the nominal part of the control input and $u_r(t)\in\mathbb{R}$ is a robustifying term that will be selected to dominate the closed-loop perturbation caused by the RBFNN approximation error. Under these considerations, we select the following control and adaptation laws:\begin{align}
    u_n&={-}\Big[ f_n(x){+}Z^T\hat{w}{+}\frac{\bar{x}_2^2{-}x_2^2}{\bar{x}_2^2}( s_1{-}\rho_2\dot{a}_1){+}k_2z_2 {+}\hat{d}\Big]\label{eq:lyapunov-control-law} \\[2pt]
    u_r&=-\hat{\epsilon}\tanh\Big(\frac{s_2}{\phi_p}\Big) \label{eq:robustifying-term} \\[2pt]
    \dot{\hat{b}} &=\gamma_1\operatorname{Proj}\Big( \hat{b},s_2u-\sigma_m\hat{b} \Big) \label{eq:control-gain-adaptation-law} \\[2pt]
    \dot{\hat{w}} &= \gamma_2\operatorname{Proj}\Big( \hat{w},s_2Z(x)-\sigma_m\hat{w} \Big) \label{eq:rbfnn-weights-adaptation-law} \\[2pt]
    \dot{\hat{d}} &= \gamma_4\operatorname{Proj}\Big( \hat{d},s_2-\sigma_m\hat{d} \Big) \label{eq:disturbance-adaptation-law}\\[2pt]
    \dot{\hat{\epsilon}} &= \gamma_3\operatorname{Proj}\Big( \hat{\epsilon}, |s_2|-\sigma_m\hat{\epsilon} \Big) \label{eq:rbfnn-error-adaptation-law}
\end{align}where $\phi>0$ is a design parameter and $\sigma_m>0$ is a leakage term included to prevent drift of the estimates and to facilitate closed-loop analysis (see section \ref{subsec:lyap-closed-loop-analysis}). The admissible region of each estimated quantity is shaped by the projection operator:\begin{equation}
    \operatorname{Proj}(\hat{\theta},k) = \begin{cases}
        k-h, &G(\hat{\theta})>0 \ \& \ k{\cdot}\nabla G(\hat{\theta})>0 \\
        k, & \text{ otherwise}
    \end{cases} \label{eq:projection-operator}
\end{equation}where $h = \frac{\nabla G(\hat{\theta})^2}{|\nabla G(\hat{\theta})|^2}{\cdot} k{\cdot} G(\hat{\theta})$ and $G(\hat{\theta})=\frac{|\hat{\theta}|^2-\bar{\theta}^2}{\epsilon_\theta\cdot \bar{\theta}^2}$ with $\bar{\theta}>0$ denoting the absolute upper bound of the estimated quantity and $\epsilon_\theta>0$ a parameter that specifies the boundary tolerance. Notice that based on \eqref{eq:projection-operator}, the norm of the estimated vector $\hat{\theta}$ will be maintained strictly inside the closed set defined as:\begin{equation}
    \mathcal{S}_{\hat{\theta}} := \Big\{\hat{\theta}\in\mathbb{R} \ : \ ||\hat{\theta}||\leq\overline{\theta}_{max}\cdot\sqrt{1+\epsilon_\theta}=\bar{\Theta}\Big\} \label{eq:projection-closed-set}
\end{equation}An important property of \eqref{eq:projection-operator} that will be used in the subsequent closed-loop analysis is the following:\begin{equation}
    \tilde{\theta}^T\Big(\operatorname{Proj}(\hat{\theta},k)-k\Big)\leq0 \label{eq:projection-inequality}
\end{equation}These arguments hold also for the scalar case, where $\hat{\theta}\in\mathbb{R}$. By virtue of \eqref{eq:system-dynamics}, it is necessary to guarantee $\hat{b}>0$ and the same holds for $\hat{\epsilon}>0$, as the latter acts like a time-varying gain of the robustifying control term \eqref{eq:robustifying-term}. Consequently, the admissible regions of these two quantities are defined as\begin{align}
    \hat{b}&\in[b_{min} \ b_{max}] & \hat{\epsilon}&\in[\epsilon_{min} \ \epsilon_{max}] \label{eq:control-gain-epsilon-estimates-intervals}
\end{align}where $0<b_{min},\epsilon_{min}<b_{max},\epsilon_{max}$. To ensure \eqref{eq:control-gain-epsilon-estimates-intervals}, the adaptation laws \eqref{eq:control-gain-adaptation-law} and \eqref{eq:rbfnn-error-adaptation-law} are deployed with $\hat{b}-(b_{max}+b_{min})/2$ and $\hat{\epsilon}-(\epsilon_{max}+\epsilon_{min})/2$, respectively. These constitute a shifting of coordinates by a constant factor, the derivative of which is zero, thereby preserving the aforementioned properties of \eqref{eq:projection-operator} without compromising the subsequent analysis.

Substituting \eqref{eq:lyapunov-control-law} - \eqref{eq:rbfnn-error-adaptation-law} into \eqref{eq:second-step-lyap-der} yields\begin{align}
    &\dot{V}_c = -W(x_1,r)-k_2\frac{\bar{x}_2^2z_2^2}{\bar{x}_2^2-x_2^2}-\hat{\epsilon}s_2\tanh\Big( \frac{s_2}{\phi_p} \Big) + s_2\epsilon(x) \notag \\[3pt]
    &+\tilde{b}\big(\operatorname{Proj}(\hat{b},s_2u-\sigma_m\hat{b})-s_2u\big) \label{eq:second-step-final-lyap-der} \\[3pt]
    &+ \tilde{w}^T\big( \operatorname{Proj}(\hat{w},s_2Z(x)-\sigma_m\hat{w}) -s_2Z(x)\big) \notag \\[3pt]
    &+\tilde{d}\big(\operatorname{Proj}(\hat{d},s_2-\sigma_m\hat{d} )-s_2) +\tilde{\epsilon}\big(\operatorname{Proj}(\hat{\epsilon},|s_2|-\sigma_m\hat{\epsilon})\big) \notag
\end{align}

\textit{Remark 2}: Note that computation of \eqref{eq:lyapunov-control-law} - \eqref{eq:rbfnn-error-adaptation-law} involves computation of $\rho_1,\rho_2$. However, it is evident that both $\rho_1,\rho_2$ have closed-form solutions $\rho_1=\frac{\bar{x}_1}{2z_1}\ln\big( \frac{(\bar{x}_1+x_1)(\bar{x}_1-r)}{(\bar{x}_1-x_1)(\bar{x}_1+r)} \big)$ if $z_1\neq0$ and $\rho_1=\frac{\bar{x}_1^2}{\bar{x}_1^2-r^2}$ if $z_1=0$. Similarly, for $\rho_2$. 

\subsection{Closed-loop Analysis}
\label{subsec:lyap-closed-loop-analysis}
\begin{theorem} Consider \eqref{eq:system-dynamics} satisfying assumption 1--3 in closed-loop under the control and adaptation laws \eqref{eq:lyapunov-control-law} - \eqref{eq:rbfnn-error-adaptation-law} and a reference trajectory $r(t):[0 \ \infty)\rightarrow\mathbb{R}$ satisfying assumption 4. Provided that the set of initial conditions $x(0)$ satisfies $|x_i|<\bar{x}_i,i=\{1,2\}$, then the following conditions hold:\begin{enumerate}[leftmargin=*]
    \item The closed-loop system is uniformly bounded:\begin{equation}
        V_c(t) \leq \Big(V_c(0) - \frac{\mu_o}{\lambda}\Big)e^{-\lambda t}+\frac{\mu_o}{\lambda}, \forall t\geq 0 \label{eq:theorem-1-first-argument}
    \end{equation}where the convergence rate $\lambda>0$ and the constant $\mu_o>0$ depend on the controller parameters. Inequality \eqref{eq:theorem-1-first-argument} implies that for any $\eta>0$, there exists a time $T_\eta\geq 0$ s.t. for all $t\geq T_\eta$ $V_c(t)\leq \frac{\mu_o}{\lambda} + \eta$.
    \item The tracking error metrics $z_i,i=\{1,2\}$ converge to a ball that contains the origin and is defined for $c_{1i}>0$ as:\begin{equation}
        \mathcal{Z}_i:=\Bigg\{z_i\in\mathbb{R} \ : \ |z_i(t)|\leq \sqrt{\frac{1}{c_{1i}}\Big( \frac{\mu_o}{\lambda}+\eta \Big)}\Bigg\} \label{eq:theorem-1-second-argument}
    \end{equation}
    \item The state constraints are satisfied, that is:\begin{equation}
        \big(x_1(t),x_2(t)\big)\in(-\bar{x}_1 \ \bar{x}_1)\times (-\bar{x}_2 \ \bar{x}_2), \forall t\geq 0 \label{eq:theorem-1-third-argument}
    \end{equation}
    \item All closed-loop signals remain bounded $\forall t\geq 0$.
\end{enumerate}\end{theorem}

\begin{proof} Let $[0, T_{\operatorname{max}})$ denote the maximal interval on which the closed-loop solution exists and remains in the admissible set. To begin, we first utilize \eqref{eq:projection-inequality} and get the following:\begin{align}
    &\tilde{b}\big(\operatorname{Proj}(\hat{b},s_2u-\sigma_m\hat{b})-(s_2u-\sigma_m\hat{b})\big) \leq 0 \Rightarrow \notag \\[2pt]
    \Rightarrow &\tilde{b}\big(\operatorname{Proj}(\hat{b},s_2u-\sigma_m\hat{b})-s_2u\big) \leq -\sigma_m\tilde{b}\hat{b} \label{eq:proof-1-proje-ineq-utilization}
\end{align}Through identical procedures, we also obtain:\begin{align}
    &\tilde{w}^T\big(\operatorname{Proj}(\hat{w},s_2Z(x)-\sigma_m\hat{w})-s_2Z(x)\big) \leq -\sigma_m\tilde{w}^T\hat{w} \label{eq:proof-1-proj-ineq-utilization-2} \\[3pt]
    &\tilde{d}\big(\operatorname{Proj}(\hat{d},s_2-\sigma_m\hat{d})-s_2\big)\leq -\sigma_m\tilde{d}\hat{d} \\[3pt]
    &\tilde{\epsilon}\big(\operatorname{Proj}(\hat{\epsilon},|s_2|-\sigma_m\hat{\epsilon})\big) \leq \tilde{\epsilon}|s_2|-\sigma_m\tilde{\epsilon} \hat{\epsilon} \label{eq:proof-1-proj-ineq-utilization-3}
\end{align}Substituting \eqref{eq:proof-1-proje-ineq-utilization} - \eqref{eq:proof-1-proj-ineq-utilization-3} into \eqref{eq:second-step-final-lyap-der} yields:\begin{align}
    \dot{V}_c\leq &-W(x_1,r)-k_2\frac{\bar{x}_2^2z_2^2}{\bar{x}_2^2-x_2^2}-\hat{\epsilon}s_2\tanh\Big( \frac{s_2}{\phi} \Big) + s_2\epsilon(x) \notag \\[2pt]
    &-\sigma_m\tilde{b}\hat{b}-\sigma_m\tilde{w}^T\hat{w}+\tilde{\epsilon}|s_2| -\sigma_m\tilde{d}\hat{d} -\sigma_m\tilde{\epsilon}\hat{\epsilon} \label{eq:proof-1-first-lyap-der}
\end{align}To proceed, we use, for $s\in\mathbb{R}$ and $\phi_p>0$, inequality $|s|-s\tanh(s/\phi_p)\leq c_{\text{tanh}}\phi_p=0.2785\phi_p$, define $T_\epsilon=s_2\epsilon(x)-\hat{\epsilon}s_2\tanh(s_2/\phi_p)$ and recall that $|\epsilon(x)|\leq \bar{\epsilon}$:\begin{align}
    &T_\epsilon\leq-(\hat{\epsilon}-\bar{\epsilon})|s_2|+\hat{\epsilon}\big(|s_2|{-}s_2\tanh(s_2/\phi_p)\big) \notag \Rightarrow \\[2pt]
    \Rightarrow& T_\epsilon\leq-\tilde{\epsilon}|s_2|+\hat{\epsilon}\big(|s_2|-s_2\tanh(s_2/\phi_p)\big) \Rightarrow \label{eq:approximation-error-terms-bound} \\[2pt]
    \Rightarrow&T_\epsilon\leq-\tilde{\epsilon}|s_2|+0.2785\phi_p\hat{\epsilon} \notag
\end{align}To proceed, notice that the following holds:\begin{equation}
    -\sigma_m\tilde{b}\hat{b}{=}{-}\sigma_m\tilde{b}(\tilde{b}+b)={-}\sigma_m\tilde{b}^2{-}\sigma_m\tilde{b}b{\leq}{-}\sigma_m|\tilde{b}|^2{+}\sigma_m|\tilde{b}||b| \notag
\end{equation}
Applying Young's inequality to the second term:\begin{equation}
    |\tilde{b}||b|\leq\frac{1}{2}|\tilde{b}|^2+\frac{1}{2}|b|^2
\end{equation} we get the following bound\begin{equation}
    -\sigma_m\tilde{b}\hat{b}\leq-\frac{\sigma_m}{2}|\tilde{b}|^2+\frac{\sigma_m}{2}|b|^2 \label{eq:proof-1-control-gain-term-bound}
\end{equation}Following similar procedures, we obtain:\begin{align}
    &-\sigma_m\tilde{w}^T\hat{w}\leq-\frac{\sigma_m}{2}||\tilde{w}||^2+\frac{\sigma_m}{2}||w^*||^2 \label{eq:proof-1-rbfnn-weights-bound} \\[2pt]
    &-\sigma_m\tilde{d}\hat{d}\leq-\frac{\sigma_m}{2}\tilde{d}^2+\frac{\sigma_m}{2}d^2 \label{eq:proof-1-disturbance-term-bound} \\[2pt]
    &-\sigma_m\tilde{\epsilon}\hat{\epsilon}\leq-\frac{\sigma_m}{2}|\tilde{\epsilon}|^2+\frac{\sigma_m}{2}\bar{\epsilon}^2 \label{eq:proof-1-rbfnn-error-term-bound}
\end{align}Substituting \eqref{eq:approximation-error-terms-bound}, \eqref{eq:proof-1-control-gain-term-bound} - \eqref{eq:proof-1-rbfnn-error-term-bound} into \eqref{eq:proof-1-first-lyap-der}, considering that $\hat{\epsilon}\leq\hat{\bar{\epsilon}}=\epsilon_{\text{max}}\sqrt{1+\bar{\epsilon}_\theta}$, which is ensured through the utilization of the projection operator \eqref{eq:rbfnn-error-adaptation-law} and its property \eqref{eq:projection-closed-set} and recalling assumption 3, yields:\begin{align}
    \dot{V}_c&\leq-k_1\frac{\bar{x}_1^2z_1^2}{\bar{x}_1^2-x_1^2}-k_2\frac{\bar{x}_2^2z_2^2}{\bar{x}_2^2-x_2^2}-\frac{\sigma_m}{2}\big( |\tilde{b}|^2+||\tilde{w}||^2 +\tilde{d}^2 \notag \\[2pt] &+|\tilde{\epsilon}|^2 \big) +\underbrace{0.2785\phi_p\hat{\bar{\epsilon}}+\frac{\sigma_m}{2}\big(b^2+||w^*||^2+\bar{\epsilon}^2{+}\bar{D}^2\big)}_{:=\mu_o} \label{eq:proof-1-second-lyap-der}
\end{align}To proceed, we utilize the standard iBLF comparison bounds that are valid on the admissible set $|x_i|<\bar{x}_i$ that state that there exist positive constants $c_{1i},c_{2i}$ s.t. \cite{tang2016}:\begin{equation}
    c_{1i}\frac{\bar{x}_i^2z_i^2}{\bar{x}_i^2-x_i^2}\leq V_{x_i}(z_i,\cdot)\leq c_{2i}\frac{\bar{x}_i^2z_i^2}{\bar{x}_i^2-x_i^2} , \ i=\{1,2\}\label{eq:proof-1-lyapunov-functions-bounds}
\end{equation}Based on the upper bound of \eqref{eq:proof-1-lyapunov-functions-bounds}, we obtain\begin{equation}
    \frac{\bar{x}_i^2z_i^2}{\bar{x}_i^2-x_i^2}\geq \frac{1}{c_{2i}}V_{x_i}(z_i,\cdot) , \ i=\{1,2\}
\end{equation}Summing over and multiplying by $k_1,k_2$, yields:\begin{equation}
    k_1\frac{\bar{x}_1^2z_1^2}{\bar{x}_1^2-x_1^2}+k_2\frac{\bar{x}_2^2z_2^2}{\bar{x}_2^2-x_2^2}\geq\lambda_0(V_{x_1}+V_{x_2}) \label{eq:proof-1-part-of-final-ineq}
\end{equation} for $\lambda_0=\min(k_1/c_{21},k_2/c_{22})$. Combining \eqref{eq:proof-1-second-lyap-der} and \eqref{eq:proof-1-part-of-final-ineq} and multiplying with the adaptation rates $\gamma_j,j=\{1,2,3\}$, we get:\begin{align}
    \dot{V}_c&\leq -\lambda_0(V_{x_1}+V_{x_2})- (\sigma_m\gamma_1)\frac{\tilde{b}^2}{2\gamma_1} -(\sigma_m\gamma_2)\frac{\tilde{w}^T\tilde{w}}{2\gamma_2}\notag \\
    &-(\sigma_m\gamma_4)\frac{\tilde{d}^2}{2\gamma_4}-(\sigma_m\gamma_3)\frac{\tilde{\epsilon}^2}{2\gamma_3} +\mu_o \label{eq:proof-1-first-to-last-lyap-der} \\
    &=-\lambda\Big( V_{x_1}+V_{x_2}+ \frac{\tilde{b}^2}{2\gamma_1}+\frac{\tilde{w}^T\tilde{w}}{2\gamma_2}+\frac{\tilde{\epsilon}^2}{2\gamma_3} + \frac{\tilde{d}^2}{2\gamma_4} \Big)+\mu_o \notag
\end{align}for $\lambda:=\text{min}(\lambda_0,\sigma_m\gamma_i)>0,i=1,\dots,4$. By virtue of \eqref{eq:proof-1-first-to-last-lyap-der}, we get $\dot{V}_c(t)\leq-\lambda V_c(t)+\mu_o$, which implies that \eqref{eq:theorem-1-first-argument} holds. Since $V_c(t)$ is finite on $[0,T_{\max})$ and the iBLFs satisfy $V_{x_i}\to\infty$ as $|x_i|\to\bar{x}_i$, the state cannot reach the constraint boundary in finite time; hence $T_{\max}=\infty$. The latter also implies that \eqref{eq:composite-lyapunov-function} remains bounded, which in turn establishes $V_{x_i}\in\mathcal{L}_\infty,i=\{1,2\}$ and the state constraints are satisfied. Furthermore, boundedness of \eqref{eq:composite-lyapunov-function} implies $x,\tilde{b},\tilde{w},\tilde{\epsilon},\tilde{d}\in\mathcal{L}_\infty$, and thus, it is concluded that all closed-loop signals remain bounded. By virtue of the lower bound of \eqref{eq:proof-1-lyapunov-functions-bounds}:\begin{equation}
    c_{1i}\frac{\bar{x}_i^2z_i^2}{\bar{x}_i^2-x_i^2}\leq V_{x_i}\leq V_c\leq\frac{\mu_o}{\lambda}+\eta
\end{equation}Note that since $|x_i|<\bar{x}_i\Rightarrow\frac{\bar{x}_i^2}{\bar{x}_i^2-x_i^2}\geq 1$. The latter implies\begin{equation*}
    z_i^2\leq\frac{\bar{x}_i^2z_i^2}{\bar{x}_i^2-x_i^2}\leq \frac{1}{c_{1i}}\Big( \frac{\mu_o}{\lambda}+\eta \Big) \Rightarrow |z_i(t)|\leq\sqrt{\frac{1}{c_{1i}}\Big( \frac{\mu_o}{\lambda+\eta} \Big)}
\end{equation*}thereby proving 2) of theorem 1.\end{proof}

\textit{Remark 3}: Based on the analysis, the condition $|a_1|<\bar{x}_2$ must hold and to this end $k_1$ is chosen to satisfy $|a_1(0)|<\bar{x}_2$, which based on theorem 1 establishes $|a_1(t)|<\bar{x}_2,\forall t\geq 0$. For higher-order systems, a feasibility check should be performed based on an offline constrained optimization problem to properly select the gains $k_i,i=1,\dots,n-1$ \cite{ktee2012,tang2016}.



\textit{Remark 4}: Although the analysis is presented for system \eqref{eq:system-dynamics}, the results extend directly to systems consisting of several subsystems of the same structure for $i=1,\dots,m$: $\dot{x}_{i,1} = x_{i,2}, \ 
\dot{x}_{i,2} = f_{n,i}(x) + b_i u_i + F_i(x) + d_i(t)$ where the overall state vector is defined as $x = [x_{1,1}, x_{1,2}, x_{2,1}, x_{2,2}, \dots, x_{m,1}, x_{m,2}]^T$. The design can be applied independently to each subsystem, resulting in a composite Lyapunov function obtained as the sum of the individual Lyapunov functions. This structure naturally arises in the quadcopter attitude dynamics considered in section~\ref{sec:exp-sim-scenarios}.

\textit{Remark 5}: Theorem 1 implicitly assumes constant disturbances. However, the result extends to piecewise-constant disturbances provided that a dwell time $\tau_d>0$ between jumps exists. At each discontinuity, \eqref{eq:composite-lyapunov-function} exhibits a finite jump associated with the instantaneous $\tilde{d}$, but remains bounded. Therefore, for finitely many jumps, boundedness of \eqref{eq:composite-lyapunov-function} is preserved across jumps and after the final jump, properties of theorem 1 are ultimately recovered.

\section{RALMPC Controller}
\label{sec:ralmpc-design}
\subsection{Sampled-data Implementation}
\label{subsec:sampled-data-implementation}
In practice, the control and adaptation laws \eqref{eq:lyapunov-control-law}--\eqref{eq:rbfnn-error-adaptation-law} are implemented in sample-and-hold form with sampling period $T_s>0$. Since the iBLFs keep the state within the compact admissible set $|x_i|<\bar{x}_i$, $i=1,2$, the closed-loop vector field is locally Lipschitz and all signals remain bounded. The sample-and-hold input can be expressed as the continuous-time law plus a mismatch $\Delta_u(t)$, which satisfies $|\Delta_u(t)|\le cT_s$ for $c>0$. Hence, the sampling effect appears as an additional bounded perturbation in \eqref{eq:second-step-final-lyap-der}, preserving $\dot{V}_c(t)\le -\lambda V_c(t)+\mu_{o,SH}$ with $\mu_{o,SH}=\mu_o+\mathcal{O}(T_s)$. Therefore, uniform boundedness is retained for sufficiently small $T_s$ \cite{liberzon2003,hetel2017}.

\subsection{Contractive Constraint}
\label{subsec:ralmpc-contractive-constraint}
The feedback controller under RALMPC is:\begin{equation}
    u(t) = \hat{b}(t)^{-1}\cdot\big[u_{mpc}(t) + u_r(t)\big] \label{eq:mpc-feedback-controller}
\end{equation}Based on \eqref{eq:mpc-feedback-controller}, \eqref{eq:second-step-lyap-der} becomes:\begin{align}
    &\dot{V}_{c,mpc} ={-}W(x_1,r)+z_2\Big[s_1+\frac{\bar{x}_2^2}{\bar{x}_2^2-x_2^2}\big[f_n(x) + u_{mpc}+u_r \notag \\
    &+Z(x)^T\hat{w}+\hat{d}\big] -\rho_2\dot{a}_1\Big] + \tilde{b}\big(\gamma_1^{-1}\dot{\hat{b}}-s_2u\big) \label{eq:mpc-lyap-der-first} \\
    &+ \tilde{w}^T\big(\gamma_2^{-1}\dot{\hat{w}} - s_2Z(x)\big) + \tilde{d}(\gamma_4^{-1}\dot{\hat{d}}-s_2)+\gamma_3^{-1}\tilde{\epsilon}\dot{\hat{\epsilon}} + s_2\epsilon(x)\notag
\end{align}Substituting \eqref{eq:robustifying-term} - \eqref{eq:rbfnn-error-adaptation-law} into \eqref{eq:mpc-lyap-der-first} yields:\begin{align}
    &\dot{V}_{c,mpc}-W(x_1,r)+z_2\Big[ s_1 + \frac{\bar{x}_2^2}{\bar{x}_2^2-x_2^2}\big[ f_n(x)+u_{mpc} \notag \\
    &+Z(x)^T\hat{w} +\hat{d} \big] -\rho_2\dot{a}_1 \Big] -\hat{\epsilon}s_2\tanh\Big( \frac{s_2}{\phi} \Big) + s_2\epsilon(x) \notag \\
    &+\tilde{b}\big(\operatorname{Proj}(\hat{b},s_2u-\sigma_m\hat{b})-s_2u\big) \label{eq:mpc-second-lyap-der} \\
    &+ \tilde{w}^T\big( \operatorname{Proj}(\hat{w},s_2Z(x)-\sigma_m\hat{w}) -s_2Z(x)\big) \notag \\
    &+\tilde{d}\big(\operatorname{Proj}(\hat{d},s_2-\sigma_m\hat{d} )-s_2) +\tilde{\epsilon}\big(\operatorname{Proj}(\hat{\epsilon},|s_2|-\sigma_m\hat{\epsilon})\big) \notag
\end{align}The contractive constraint at time $t$ is formed as:\begin{equation}
    \dot{V}_{c,mpc}(t) \leq \dot{V}_c(t) \label{eq:general-formulation-of-contr-con}
\end{equation}By virtue of \eqref{eq:second-step-final-lyap-der}, \eqref{eq:mpc-second-lyap-der} and \eqref{eq:general-formulation-of-contr-con}, the contractive constraint, after eliminating common terms, is constructed as follows:\begin{align}
    z_2\cdot\Big[ &s_1 + \frac{\bar{x}_2^2}{\bar{x}_2^2-x_2^2}\cdot \big[ f_n(x)+u_{mpc}+Z(x)^T\hat{w} +\hat{d} \big] \notag \\
    &-\rho_2\dot{a}_1 \Big] \leq -k_2\frac{\bar{x}_2^2}{\bar{x}_2^2-x_2^2}\cdot z_2^2 \label{eq:final-contractive-constraint-formulation}
\end{align}

\subsection{Relaxed Polynomial Barrier Function}
\label{eq:subsec:rpbf-dead-zone}

As stated in Section~\ref{subsec:control-prob-form}, RALMPC must account for input limits without compromising OCP feasibility. Since the contractive constraint based on \eqref{eq:lyapunov-control-law} does not explicitly account for hard control bounds, such constraints cannot be imposed directly in the OCP. Instead, inspired by \cite{feller2017}, we introduce a relaxed polynomial barrier function with dead-zone (RPBF). Compared with \cite{feller2017}, the RPBF strongly penalizes values outside the admissible set while leaving a dead-zone unpenalized, which is crucial because the varying input equilibrium is unavailable under uncertainties and disturbances.

\begin{figure}[btp]
\centerline{\includegraphics[width=\columnwidth]{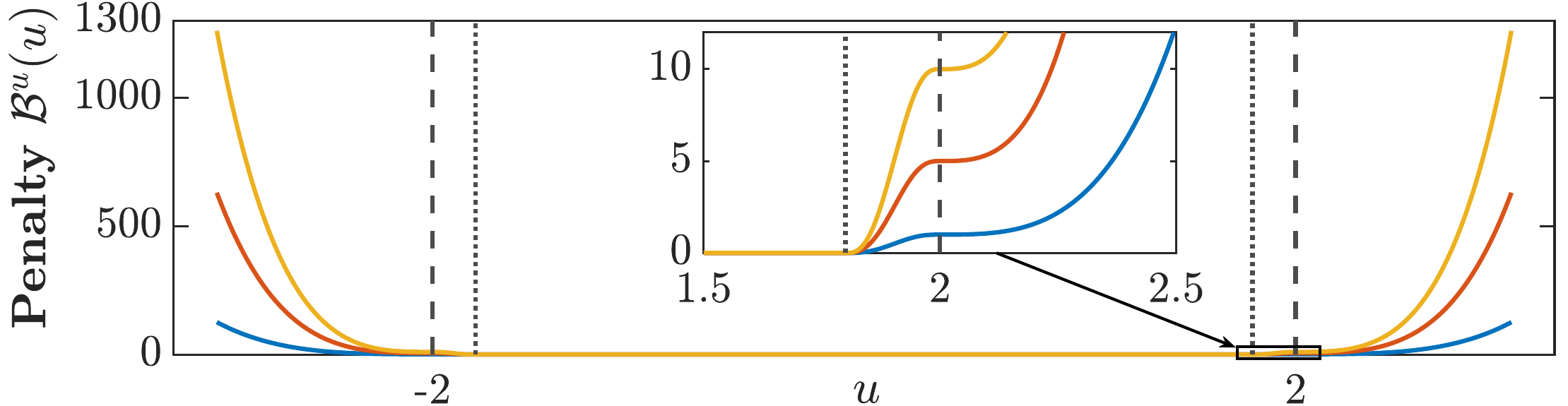}}
\caption{RPBF with $|u_{max}|=2,u_\epsilon=0.2$: $\sigma_u=1,c=1$ (solid blue) - $\sigma_u=5,c=5$ (solid red) - $\sigma_u=10,c=10$ (solid yellow) - bounds (black dashed) - boundary layer interval (black dotted)}
\label{fig:dead-zone-rpbf}
\vspace{-0.5cm}
\end{figure}

To avoid non-smoothness at $u=0$, we define the RPBF as a piecewise polynomial function of $s=u^2\Rightarrow s>0$, thus yielding an even penalty at the points $u$ and $-u$. Regarding the dead-zone and boundary layer, let $q=u_{max}-u_\epsilon, s_q=q^2$ and $s_b=u_{max}^2$, so that the dead-zone corresponds to $s\leq s_q$ and the boundary to $s=s_b$; parameter $u_\epsilon>0$ defines the interval between the dead-zone and $u_{max}$, inside which $u$ starts getting penalized. For the boundary layer interval $s\in(s_q \ s_b)$, we introduce the normalized coordinate\begin{equation}
    \phi(s)=\frac{s-s_q}{s_b-s_q}\in(0 \ 1) \label{eq:rpbf-boundary-interval}
\end{equation}and utilize a quintic polynomial of the form:\begin{equation}
    h(\phi)=10\phi^3-15\phi^4+6\phi^5 \label{eq:rpbf-quintic-polynomial}
\end{equation}which satisfies the conditions $h(0)=h^{'}(0)=h^{''}(0)=h^{'}(1)=h^{''}(1)=0,h(1) = 1$. These ensure that the first and second order derivatives vanish at the splice points, thus guaranteeing that the RPBF is a $\mathcal{C}^2$ function. For $s\geq s_b \ (|u|\geq u_{max})$, we extend the penalty with a cubic polynomial in $\xi=s-s_b$, namely $\xi^3$, which satisfies:\begin{align}
    &\xi^3\Big|_{\xi=0}=\frac{d}{ds}\Big|_{\xi=0}=\frac{d^2}{ds^2}\Big|_{\xi=0} = 0
\end{align}This choice is deliberate to match value, slope, and curvature at the boundary point $s=s_b$. Consequently, the RPBF is defined as $\mathcal{B}^u(u)=\Phi(u^2)$ with:\begin{equation}\label{eq:deadzone_barrier}
\Phi(s)\triangleq
\begin{cases}
0, & s\le s_q,\\[2pt]
\sigma_u\,h\!\left(\phi\right), & s_q<s<s_b,\\[2pt]
\sigma_u + c\,(s-s_b)^3, & s\ge s_b,
\end{cases}
\end{equation}
where \(\sigma_u>0\) is the penalty level reached at the boundary \(|u|=u_{\max}\),
and \(c>0\) controls how aggressively violations are punished outside the admissible set. A graphical illustration of \eqref{eq:deadzone_barrier} is presented in Fig. \ref{fig:dead-zone-rpbf}.

\subsection{Optimal Control Problem}
\label{subsec:ralmpc-ocp}
Under the aforementioned, the finite horizon OCP at time $t$ with initial conditions $x_0=x(t)$ is formulated as follows:\begin{subequations}
\begin{align}&\min\limits_{u_{\tau|t}}\int\limits_{t}^{t+N_p\cdot T_s}\Big( \big||\hat{x}_{1,\tau|t}-r_{\tau|t}|\big|^2_{Q_1} + \big||\hat{z}_{2,\tau|t}|\big|^2_{Q_2}+ \mathcal{B}^u_{\tau|t} \notag \\[2pt]
    &\qquad \qquad \quad + \big||\Delta u_{\tau|t}|\big|^2_{R_1}\Big)d\tau \label{eq:ocp-cost-function} \\[2pt] 
    \text{s.t. } 
    &\dot{\hat{x}}_{1,\tau|t}=\hat{x}_{2,\tau|t} \label{eq:ocp-x1-prediction-model}\\[2pt]
    &\dot{\hat{x}}_{2,\tau|t}=f_n(\hat{x}_{\tau|t})+\hat{b}_{t}{\cdot }u_{\tau|t}+Z(\hat{x}_{\tau|t})^T\cdot\hat{w}_{t} + \hat{d}_{t} \label{eq:ocp-prediction-model} \\[2pt]
    &\hat{z}_{2,\tau|t} = \hat{x}_{2,\tau|t}-\hat{a}_{1,\tau|t} \label{eq:ocp-z2-prediction-model} \\[2pt]
    &\hat{x}\in\mathcal{X}_s \label{eq:ocp-state-constraints} \\[2pt]
    &\dot{V}_{c,mpc}(t) \leq \dot{V}_{c}(t) \label{eq:ocp-contractive-constraint}
\end{align}
\end{subequations}where at time $t$, $\hat{x}_{\tau|t}$ denotes the predicted state trajectory, $u_{\tau|t}$ represents the control input trajectory returned as the OCP solution that is calculated based on \eqref{eq:mpc-feedback-controller} with constant $u_{r,\tau|t}$ along the prediction horizon, $N_p$ is the prediction horizon, $\hat{x}_{1,\tau|t},\hat{x}_{2,\tau|t},\hat{z}_{2,\tau|t}$ denote the state $x_1,x_2$ and error metric $z_2$ predicted trajectories, $\Delta u_{\tau|t}$ denotes the control input deviations between successive time-steps and $Q_1,Q_2,R_1\succ0$ are weighting matrices.

Constraint \eqref{eq:ocp-prediction-model} is the RALMPC prediction model; it is composed of $\hat{b}_t,\hat{w}_t,\hat{d}_t$ that denote the estimates of the control gain, RBFNN weight vector and external disturbance at time $t$ that are kept frozen over the prediction horizon. At each sampling instant these are updated through \eqref{eq:control-gain-adaptation-law}, \eqref{eq:rbfnn-weights-adaptation-law} and \eqref{eq:disturbance-adaptation-law}, thus leading to online model refinement. \eqref{eq:ocp-state-constraints} includes the constraints on the predicted state trajectories that are required to lie in $\mathcal{X}_s:=\{x_1,x_2\in\mathbb{R} \ : \ |x_i|\leq \bar{x}_i-s_t\}$, where $s_t>0$ corresponds to the tolerance of the applied solver, which is typically a small value $(10^{-6}$ {-} $10^{-3})$, and is included to convert $|x_i|<\bar{x}_i$ to box constraints. The latter is necessary since optimization solvers do not handle open-set constraints; however, the small value of $s_t$ does not cause conflicts with \eqref{eq:ocp-contractive-constraint}, which is the contractive constraint that allows RALMPC to inherit the properties of theorem 1. 

\begin{prop} Consider \eqref{eq:system-dynamics} and a set of initial conditions $x(0)$ at time $t=0$ satisfying $|x_i(0)|<\bar{x}_i,i=\{1,2\}$. Then, the OCP defined by \eqref{eq:ocp-cost-function} - \eqref{eq:ocp-contractive-constraint} admits recursive feasibility.
\end{prop}

\begin{proof} Provided that the set of initial conditions $x(0)$ at time $t=0$ satisfies $|x_{i}(0)|<\bar{x}_i,i=\{1,2\}$, it has been proven in theorem 1 that \eqref{eq:lyapunov-control-law} - \eqref{eq:rbfnn-error-adaptation-law} exist and are well-defined. The latter implies that $u_n(t)$ of \eqref{eq:control-summation} exists, and thus at least one solution of \eqref{eq:ocp-cost-function} - \eqref{eq:ocp-contractive-constraint} exists, implying that the feasible set of RALMPC is non-empty. The latter ensures that the constraints are satisfied, since these are included in the formulation of \eqref{eq:lyapunov-control-law}, and thus, at the next sampling time $t_1 = t+T_s$, the initial conditions of the OCP $x_0(t+T_s)$ satisfy $|x_{i,0}(t+T_s)|<\bar{x}_i$. Consequently, recursive feasibility is guaranteed.\end{proof}

\begin{theorem} Consider a set of initial conditions $x(0)$ satisfying $|x_i|<\bar{x}_i,i=\{1,2\}$, the LF defined as in \eqref{eq:composite-lyapunov-function} and \eqref{eq:system-dynamics} operating in closed-loop under the feedback law $u_{mpc}(t)$ computed as the OCP \eqref{eq:ocp-cost-function} - \eqref{eq:ocp-contractive-constraint} solution. Then, the closed-loop system inherits properties 1) - 4) of theorem 1.\end{theorem}

\begin{proof} Define the composite candidate Lyapunov function under the RALMPC controller, by virtue of \eqref{eq:composite-lyapunov-function}, as follows:\begin{equation}
    V_{c,mpc}(t)=V_{x_1}+V_{x_2}+\frac{\tilde{b}^2}{2\gamma_1}+\frac{\tilde{w}^T\tilde{w}}{2\gamma_2}+\frac{\tilde{\epsilon}^2}{2\gamma_3} +\frac{\tilde{d}^2}{2\gamma_4}\label{eq:mpc-composite-lyapunov-function}
\end{equation}Notice that by virtue of \eqref{eq:second-step-lyap-der}, \eqref{eq:second-step-final-lyap-der} and \eqref{eq:mpc-feedback-controller} and the fact that the robustifying control term \eqref{eq:robustifying-term} and adaptive laws \eqref{eq:control-gain-adaptation-law} - \eqref{eq:rbfnn-error-adaptation-law} are implemented under RALMPC, the following holds:\begin{align}
    &\dot{V}_{c,mpc} = -W(x_1,r)+z_2\Big[ s_1+\frac{\bar{x}_2^2}{\bar{x}_2^2-x_2^2}\big[f_n(x) + u_{mpc} \notag \\
    &+Z(x)^T\hat{w} + \hat{d}\big] -\rho_2\dot{a}_1 \Big]-\hat{\epsilon}s_2\tanh\Big( \frac{s_2}{\phi} \Big) + s_2\epsilon(x) \notag \\
    &+\tilde{b}\big(\operatorname{Proj}(\hat{b},s_2u-\sigma_m\hat{b})-s_2u\big) \label{eq:} \\
    &+ \tilde{w}^T\big( \operatorname{Proj}(\hat{w},s_2Z(x)-\sigma_m\hat{w}) -s_2Z(x)\big) \notag \\
    &+\tilde{d}\big(\operatorname{Proj}(\hat{d},s_2-\sigma_m\hat{d} )-s_2) +\tilde{\epsilon}\big(\operatorname{Proj}(\hat{\epsilon},|s_2|-\sigma_m\hat{\epsilon})\big) \notag
\end{align}Following the same steps as in equations \eqref{eq:proof-1-proje-ineq-utilization} - \eqref{eq:proof-1-rbfnn-error-term-bound}:\begin{align}
    &\dot{V}_{c,mpc}{\leq} {-}k_1\frac{\bar{x}_1^2z_1^2}{\bar{x}_1^2-x_1^2}{+}z_2\Big[ s_1{+}\frac{\bar{x}_2^2}{\bar{x}_2^2-x_2^2}\big[f_n(x) {+} u_{mpc} \label{eq:proof-2-first-ineq} \\[3pt]
    &+Z(x)^T\hat{w} + \hat{d}\big] -\rho_2\dot{a}_1 \Big]{-}\frac{\sigma_m}{2}\big( |\tilde{b}|^2{+}||\tilde{w}||^2 {+}\tilde{d}^2 \notag {+}|\tilde{\epsilon}|^2 \big) {+} \mu_o
\end{align}Since \eqref{eq:ocp-contractive-constraint} is satisfied, substituting \eqref{eq:final-contractive-constraint-formulation} into \eqref{eq:proof-2-first-ineq} yields:\begin{align}
    &\dot{V}_{c,mpc}\leq{-}\sum\limits_{i=1}^2k_i\frac{\bar{x}_i^2z_i^2}{\bar{x}_i^2-x_i^2}{-}\frac{\sigma_m}{2}\big( |\tilde{b}|^2{+}||\tilde{w}||^2 {+}\tilde{d}^2+|\tilde{\epsilon}|^2 \big) {+} \mu_o \notag
\end{align}Multiplying the above inequality by the adaptation rates $\gamma_i,i=\{1,2,3,4\}$, similarly to \eqref{eq:proof-1-first-to-last-lyap-der} and by virtue of \eqref{eq:mpc-composite-lyapunov-function}, yields:\begin{equation}
    \dot{V}_{c,mpc}(t)\leq-\lambda\cdot V_{c,mpc}(t)+\mu_o \label{eq:proof-2-final-ineq}
\end{equation}From \eqref{eq:proof-2-final-ineq}, it is straightforwardly deduced that properties 1) - 4) of theorem 1 are guaranteed by RALMPC.\end{proof}

\textit{Remark 6}: The fact that the contractive constraint \eqref{eq:ocp-contractive-constraint} is implemented only at the initial time $t$ of the OCP implies that it is computed using the true measurements $x(t)$. This ensures that properties of theorem 2 are formally guaranteed.

\textit{Remark 7}: Proposition 1 establishes recursive feasibility of \eqref{eq:ocp-cost-function}–\eqref{eq:ocp-contractive-constraint}. To avoid conflicts between input bounds and the contractive constraint, control limits are incorporated as relaxed constraints in the OCP cost via \eqref{eq:deadzone_barrier}. This design aligns with the BLF-based controller, which may generate large control actions near state boundaries, and accordingly, input constraints are embedded through the RPBF to preserve feasibility. Although strict satisfaction under all operating conditions formally requires hard saturation, the RPBF can be tuned to heavily penalize violations and guide the optimizer toward the admissible input set, as will be demonstrated in section \ref{sec:exp-sim-scenarios}.


\section{Simulation and Experimental Results}
\label{sec:exp-sim-scenarios}
\subsection{Quadcopter Attitude Dynamics}
\vspace{-0.1cm}
\label{subsec:quad-dynamics}
To evaluate the proposed RALMPC framework, we consider the attitude tracking problem of quadcopters. Let the orientation vector be expressed in a body-fixed coordinate frame $\{\mathcal{B}\}$ and define the state vector $x = [\phi \ p \ \theta \ q \ \psi \ r]^T$, where $\phi,\theta,\psi$ denote the roll, pitch, and yaw angles and $p,q,r$ the corresponding angular rates. Based on the latter, the attitude dynamics are described by the following equations:\begin{subequations}\label{eq:attitude-model}
    \begin{align}
        \dot{\phi} &= p &
        \dot{p} &= \frac{ (I_\theta - I_\psi)}{I_\phi}\cdot q\cdot r+ \frac{\tau_\phi}{I_\phi} + F_\phi(x) +d_\phi \\[3pt]
        \dot{\theta} &= q &
        \dot{q} &= \frac{ (I_\psi - I_\phi)}{I_\theta}\cdot p\cdot r+ \frac{\tau_\theta}{I_\theta} + F_\theta(x) + d_\theta \\[3pt]
        \dot{\psi} &= r &
        \dot{r} &=  \frac{ (I_\phi - I_\theta)}{I_\psi}\cdot p\cdot q+ \frac{\tau_\psi}{I_\psi} + F_\psi(x) + d_\psi
    \end{align}
\end{subequations}where $I_\phi,I_\theta,I_\psi>0,\tau=[\tau_\phi \ \tau_\theta \ \tau_\psi]^T$ denote the moments of inertia and torque inputs along each rotational axis, respectively, $d=[d_\phi \ d_\theta \ d_\psi]^T$ are unknown external disturbances, and $F_\phi,F_\theta,F_\psi$ are locally Lipschitz functions that represent unknown dynamics; a detailed description of terms that are lumped into $F_\phi,F_\theta,F_\psi$ is given in appendix \ref{app:quad-unknown-dynamics}.


Each attitude-axis subsystem in \eqref{eq:attitude-model} can be cast in the form of \eqref{eq:system-dynamics}; see also Remark~4. For the roll axis, we set
$b=1/I_\phi$, $F(x)=F_\phi(x)$, and 
$f_n(x)=\frac{I_\theta-I_\psi}{I_\phi}qr$,
with analogous definitions for pitch and yaw. The term $f_n(x)$ is treated as nominal because its coefficient is computed from the estimated moments of inertia and RALMPC is then used for roll, pitch, and yaw tracking, estimating the inverse inertias through \eqref{eq:control-gain-adaptation-law} and approximating $F_\phi,F_\theta,F_\psi$ with RBFNNs updated by \eqref{eq:rbfnn-weights-adaptation-law}. The operating domain is defined by:\begin{align}
    \mathcal{D}&:= (-30^{\circ} \ 30^{\circ})^3\times (-100^{\circ/s} \ 100^{\circ/s})^3 \subset \mathbb{R}^6\label{eq:quad-operating-envelope}
\end{align}

\subsection{RBFNN Formulation}
\vspace{-0.1cm}
\label{subsec:sim-exp-rbfnn-formulation}
We employ an RBFNN with an input vector that consists of the state vector $x$. We partition the domain of each of the RBFNN input variables into $3$ equal sub-intervals, thus forming a $6$-dimensional grid, the vertices of which represent the potential RBFNN hidden layer nodes. From the resulting $4^6=4096$ candidate centers, we select $150$ centers through uniform subsampling to ensure balanced coverage across all dimensions; the width of each node is computed based on the P-Nearest-Neighbor heuristic with $L=150-1=149$ neighbors for each node \cite{melagkraki2006}. Then a set of $3$ separate output weight vectors is considered to assign a distinct RBFNN model to each rotational axis; the initial values of the weights are randomly selected from a uniform distribution and set to values of the order $10^{-2}$.

\subsection{Hardware-in-the-loop Simulations}
\label{subsec:simulations}
In this section, we conduct a simulation case study to compare the proposed RALMPC controller against alternative state-of-the-art approaches both in terms of computational complexity and closed-loop performance. More specifically, we compare our approach against: 1) the Lyapunov-based controller (LBC) shaped by equations \eqref{eq:lyapunov-control-law} - \eqref{eq:rbfnn-error-adaptation-law}, 2) a Lyapunov-based MPC (LMPC) framework that is purely based on a nominal model without any adaptation, 3) a classic nonlinear MPC (NMPC) framework based on the same nominal model as LMPC with a terminal set constraint and terminal cost function without adaptation, 4) the RTMPC approach presented in \cite{kohler2021}, and 5) the RATMPC framework presented in \cite{sasfi2023}.

\subsubsection{Setup Description}
\label{subsubsec:sim-setup-description} 
In this subsection, we consider a setpoint-stabilization scenario. The initial conditions are $x=0^{6\times 1}$, and the setpoints are formulated as follows:\begin{equation*}
    r(t){=}[\phi_r(t) \ \theta_r(t) \ \psi_r(t)]^T{=}\begin{cases}
        [10^o \ 15^o \ {-}10^o]^T, & t{\in}[0 \ 8] \\[2pt]
        [{-}5^o \ 5^o \ 10^o]^T,& t{\in}(8 \ 16] \\[2pt]
        [5^o \ {-}10^o \ 25^o]^T, &t{\in}(16 \ 25]
    \end{cases}
\end{equation*}Moreover, we consider external disturbances of the form:\begin{equation*}
    d(t){=}[d_\phi \ d_\theta \ d_\psi]^T{=}\begin{cases}
        [0 \ 0 \ 0]^T, &t{\in}[0 \ 5)\\[2pt]
        [0.3 \ 0.3 \ 0.2]^T, & t{\in}[5 \ 10) \\[2pt]
        [{-}0.3 \ {-}0.3 \ 0.3]^T, & t{\in}[10 \ 15) \\[2pt]
        [{-}0.5 \ 0.2 \ 0.3]^T, & t{\in}[15 \ 20) \\[2pt]
        [0.2 \ 0.4 \ 0.8]^T, & t{\in}[20 \ 25]
    \end{cases}
\end{equation*}To ensure fairness, the sampling time for all MPC-based controllers is set to $T_s=0.02\text{sec}$, except from the LBC controller, where it is set to $T_s=0.005\text{sec}$. For the LMPC, NMPC, RTMPC, and RATMPC controllers the nominal values for the moments of inertia (MoI) are considered to be $I_\phi=I_\theta=0.0132, I_\psi=0.0194$ $[kg{\cdot}m^2]$, while the plant is simulated with $I_\phi=0.0182,I_\theta=0.0232,I_\psi=0.0294$ $[kg{\cdot}m^2]$. The former values of MoI are set as initial values of the adaptation laws \eqref{eq:control-gain-adaptation-law} for the LBC and RALMPC controllers. Moreover, for RATMPC, the employed RBFNN is considered part of its nominal model, while the RBFNN weights are set as the adaptation parameters that are being estimated online through the set-membership identification procedure; this allows to shape the attitude dynamics nominal model exactly in the form studied in \cite{sasfi2023}. The prediction horizons of NMPC, RTMPC and RATMPC are set to $N_p=15$ to ensure feasibility while maintaining the computational cost as low as possible, and the remaining parameters have been fine-tuned according to the guidelines provided in the respective works and through trial-and-error to achieve the best results; the parameters of the proposed RALMPC framework are given in Table \ref{tab:sim-ralmpc-parameters}. Notice that RALMPC requires a shorter prediction horizon, a fact that is justified by the alternative way to establish the closed-loop properties and the elimination of the terminal ingredients that are required by NMPC, RTMPC, and RATMPC; indeed using a shorter horizon for NMPC, RTMPC, and RATMPC led to degraded performance and loss of feasibility.

To ensure fairness regarding computational complexity, the MPC-based controllers are designed in a unified way; to be more specific, a single OCP is designed to account for all three rotational motions, returning the control input trajectories $\tilde{\tau}=[\tilde{\tau}_\phi \ \tilde{\tau}_\theta \ \tilde{\tau}_\psi]\in\mathbb{R}^{N_p\times 3}$ as part of its solution. The OCPs are cast into symbolic form using CasADi \cite{andersson2018} and then solved by PANOC, which is specifically tailored for embedded, nonlinear optimal control \cite{stella2017}; the interface between the solver and CasADi is set through the optimization engine (OpEn) framework \cite{sopasakis2020}. To assess real-time applicability, all simulations are conducted on a Raspberry Pi Model 4B utilizing its Broadcom BCM2711, Quad core Cortex-A72 (ARM v8) 64-bit SoC @ 1.8GHz processor and $8$GB of RAM.

\bgroup
\def\arraystretch{1.3}
\begin{table}[btp]
\centering
\begin{threeparttable}
\caption{HIL Simulation: RALMPC Parameters}
\label{tab:sim-ralmpc-parameters}
\setlength{\tabcolsep}{3pt}
\begin{tabular}
{P{12pt}P{12pt}P{12pt}P{20pt}P{70pt}P{10pt}P{10pt}P{10pt}P{17pt}}
\hline \hline
$N_p$ & $Q_1$ & $Q_2$ & $R_1$ & $\gamma_i,i=\{1,2,3,4\}$ & $k_1$ & $k_2$ & $\phi_p$ & $\sigma_m$ \\ \hline
$10$ & $20$ & $5$ & $10^{-3}$ & $\{0.1,2,10,0.75\}$ & $2.5$ & $5$ & $1$ & $10^{-3}$ \\
\hline \hline
\end{tabular}
\end{threeparttable}
\end{table}

\subsubsection{Results}
\label{subsubsec:sim-results}
To quantitatively compare the alternative control strategies, Table \ref{tab:sim-performance metrics}, depicts closed-loop evaluation metrics, namely the mean absolute error (MAE) in degrees and the $\mathcal{L}_2$-norm of the torque inputs. It is evident that RALMPC outperforms all its rivals in terms of tracking accuracy and control efficiency, as it consistently achieves lower MAE, together with reduced control effort across all rotational axes.

\bgroup
\def\arraystretch{1.4}
\begin{table}[b]
\vspace{-0.5cm}
\centering
\begin{threeparttable}
\caption{HIL Simulations: Performance Metrics}
\label{tab:sim-performance metrics}
\setlength{\tabcolsep}{4pt}
\begin{tabular}{P{40pt}P{25pt}P{25pt}P{25pt}P{25pt}P{25pt}P{25pt}}
\hline \hline
\textbf{Controller} & $\text{MAE}_{\phi}$ & $\text{MAE}_{\theta}$ & $\text{MAE}_{\psi}$ & $||\tau_\phi||_{2}$ & $||\tau_\theta||_2$ & $||\tau_\psi||_2$ \\[2pt] \hline \hline
\textbf{RALMPC} & $0.5732$ & $0.6588$ & $0.7388$ & $0.0748$ & $0.0963$ & $0.1674$ \\ 
\textbf{RATMPC} & $0.6987$ & $0.6977$ & $0.8719$ & $0.0806$ & $0.1029$ & $0.1681$ \\
\textbf{LBC} & $0.5845$ & $0.7975$ & $0.8334$ & $0.0778$ & $0.1083$ & $0.1829$ \\
\textbf{RTMPC} & $0.6867$ & $0.7781$ & $0.8882$ & $0.1028$ & $0.1401$ & $0.2082$ \\
\textbf{LMPC} & $0.7714$ & $0.9068$ & $1.0612$ & $0.1232$ & $0.1329$ & $0.1992$ \\
\textbf{NMPC} & $0.7962$ & $0.9238$ & $1.0628$ & $0.1121$ & $0.1341$ & $0.2098$ \\
\hline \hline
\end{tabular}
\end{threeparttable}
\end{table}

To visualize these performance gains, Figs. \ref{fig:sim-roll-axis} - \ref{fig:sim-yaw-axis} depict the closed-loop responses for the three rotational axes under the considered control strategies, together with the corresponding torque signals. It is evident that RALMPC achieves the best balance among all strategies w.r.t. steady-state accuracy and quick transient response without oscillations or overshoot. In contrast, while RATMPC exhibits a quicker transient response in some cases, this comes at the expense of overshoot—for example, during the second step change in the roll axis and the first step change in pitch motion; moreover, compared to RALMPC it exhibits larger steady-state error across all axes. RTMPC, on the other hand, demonstrates an overall slower response compared to RALMPC with increased control effort. However, RTMPC yields slightly better steady-state accuracy because its robust nature results in minor fluctuations during disturbance injection, whereas RALMPC exhibits slightly more fluctuations before settling back to steady-state; nonetheless, this is expected as in RALMPC, a quick transient exists during disturbance injection due to adaptation. Finally, LBC exhibits pronounced oscillations before reaching steady-state, while LMPC and NMPC demonstrate the worst performance with the larger steady-state error due to the lack of adaptive and robust mechanisms.

\begin{figure}[htbp]
    \centerline{\includegraphics[width=\columnwidth]{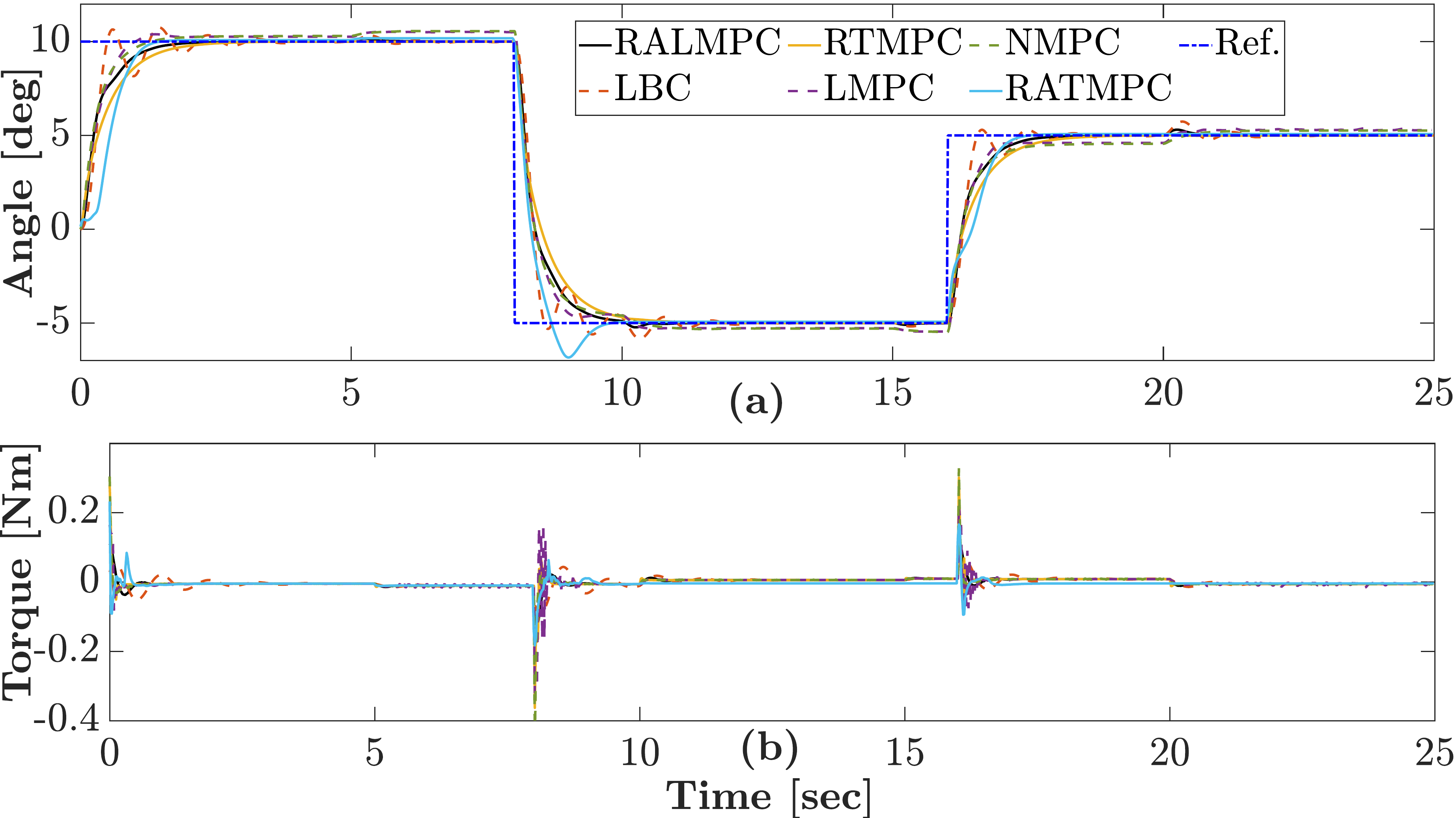}}
    \caption{Roll axis: (a) Angle - (b) Torque}
    \label{fig:sim-roll-axis}

    \vspace{0.5cm}

    \centerline{\includegraphics[width=\columnwidth]{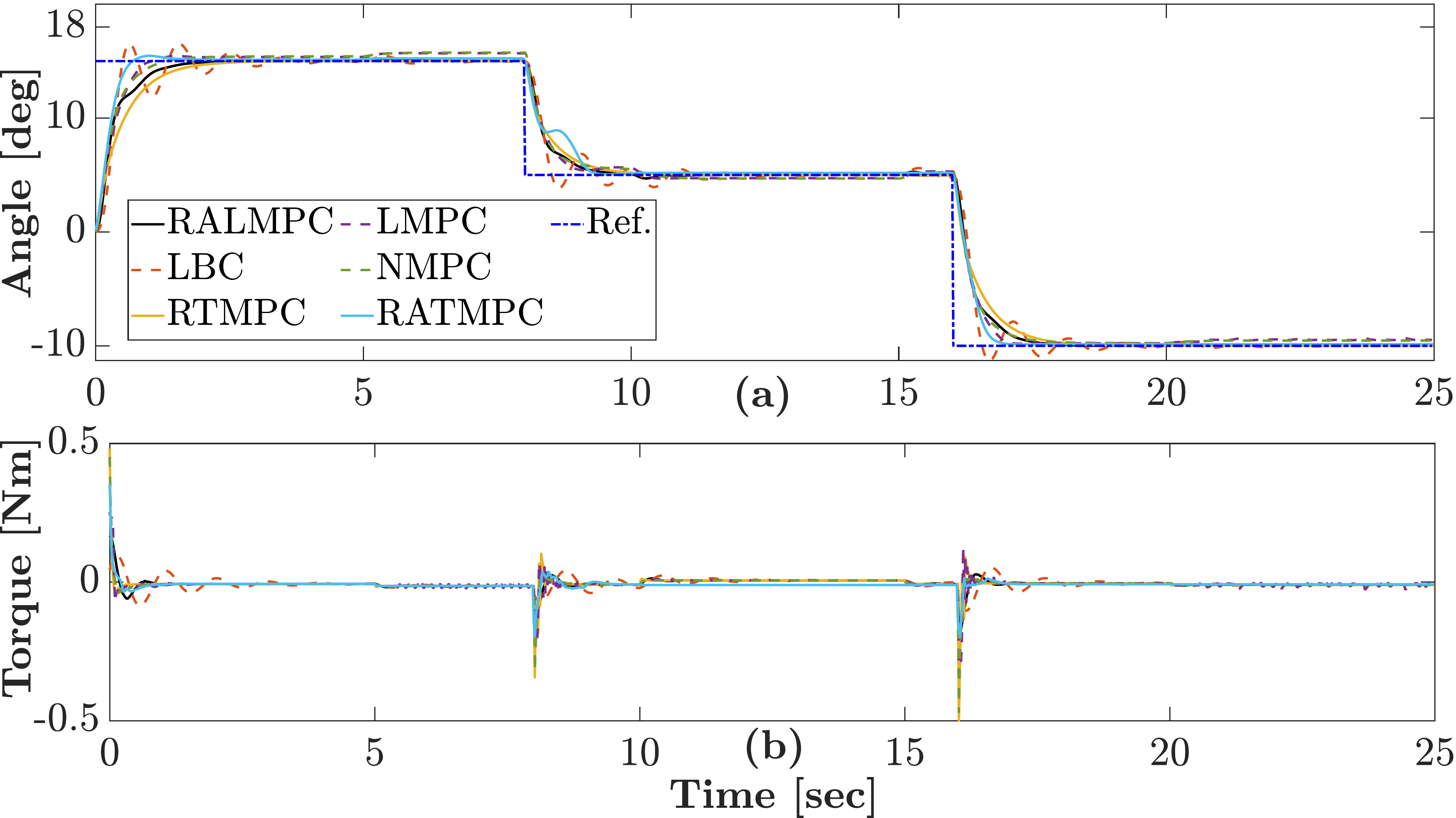}}
    \caption{Pitch axis: (a) Angle - (b) Torque}
    \label{fig:sim-pitch-axis}

    \vspace{0.5cm}

    \centerline{\includegraphics[width=\columnwidth]{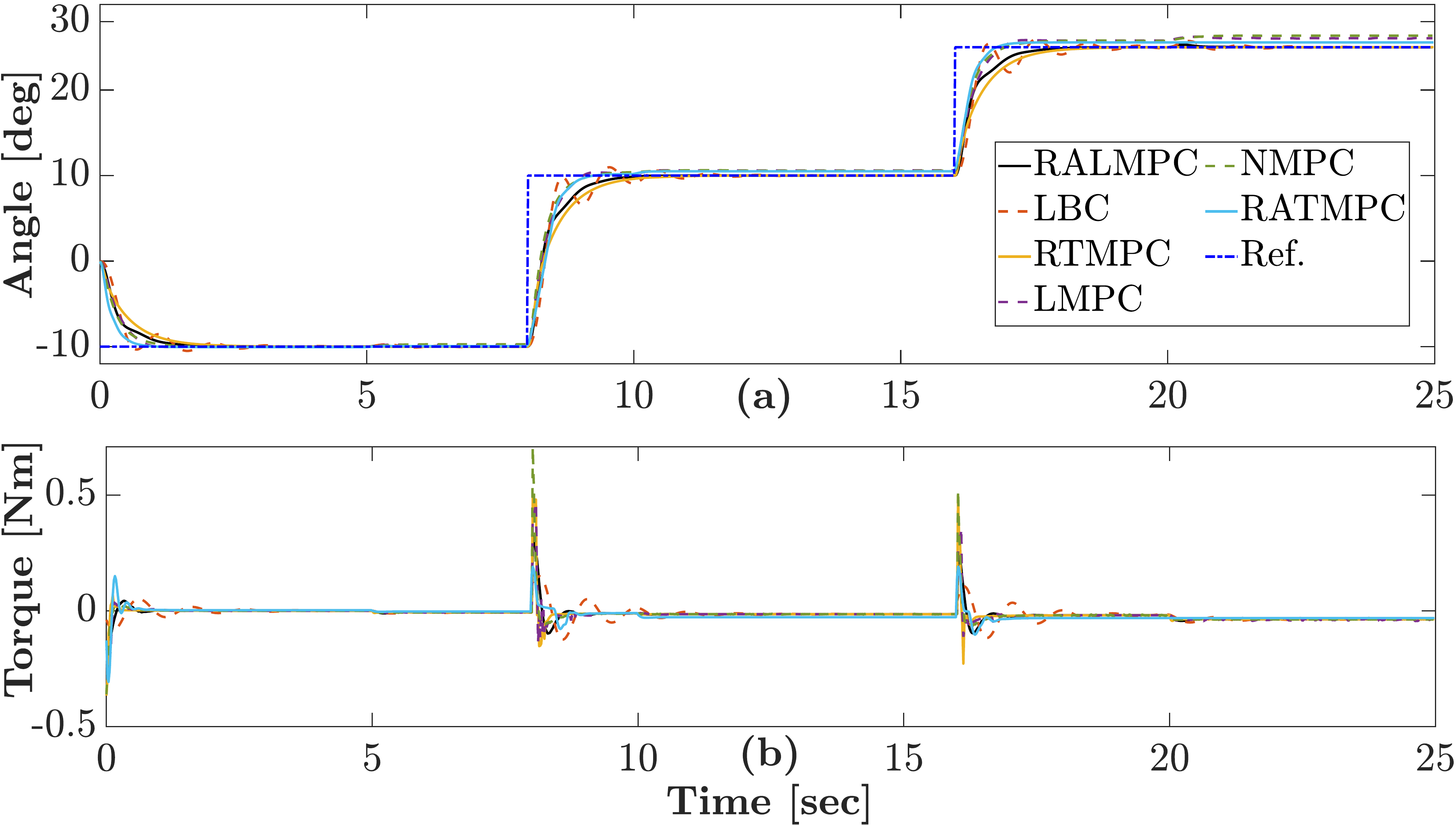}}
    \caption{Yaw axis: (a) Angle - (b) Torque}
    \label{fig:sim-yaw-axis}

    \vspace{-0.5cm}
\end{figure}




To assess the computational burden of each architecture, in Table \ref{tab:sim-computational-metrics}, we present the average $(t_{avg})$, maximum $(t_{max})$ and minimum $(t_{min})$ computational times in milliseconds required to solve the underlying OCP for each controller. It is evident that RALMPC requires significantly lower times compared to RTMPC and RATMPC, a fact that is thoroughly justified by the lack of terminal ingredients and invariant tubes; note also that RTMPC and RATMPC violate the selected sampling time $T_s$. The significant computational overhead induced by the terminal ingredients is further highlighted by the fact that even NMPC presents higher $t_{max}$ than RALMPC. As expected, LMPC is the only architecture that presents lower computational times than RALMPC, albeit at the cost of clearly inferior control performance.

\bgroup
\def\arraystretch{1.4}
\begin{table}[btp]
\centering
\begin{threeparttable}
\caption{HIL Simulations: Computational Metrics in milliseconds}
\label{tab:sim-computational-metrics}
\setlength{\tabcolsep}{4pt}
\begin{tabular}{P{25pt}P{40pt}P{35pt}P{40pt}P{30pt}P{30pt}P{30pt}}
\hline \hline
\textbf{Metric} & \textbf{RALMPC} & \textbf{RTMPC} & \textbf{RATMPC} & \textbf{LMPC} & \textbf{NMPC}\\[2pt] \hline \hline
$t_{avg}$ & $0.4252$ & $1.2143$ & $2.0562$ & $0.3142$ & $0.3841$ \\
$t_{max}$ & $1.1638$ & $54.8533$ & $82.3561$ & $1.0655$ & $2.1683$ \\
$t_{min}$ & $0.2623$ & $0.2994$ & $0.3723$ & $0.2034$ & $0.1919$ \\
\hline \hline
\end{tabular}
\end{threeparttable}
\vspace{-0.25cm}
\end{table}

\subsection{Experiment - Case I}
\label{subsec:experimental-results}
\subsubsection{Setup Description}
\label{subsubsec:exp-setup-description}
We experimentally validated the proposed RALMPC controller on a custom-built quadcopter platform depicted in Fig. \ref{fig:quadcopter}. The quadcopter is built on an F450 frame and equipped with $1100$kV brushless outrunner motors paired with $8045$ propellers and $30$A electronic speed controllers (ESCs), powered by a $2700$mAh $30$C LiPo battery. The onboard avionics include a Navio2 autopilot featuring dual IMUs, each providing triaxial gyroscope, accelerometer, and magnetometer measurements, interfaced with a Raspberry Pi 4B for onboard computation. Sensor measurements are processed through the Madgwick filter \cite{madgwick10} for attitude estimation, while gyroscope data initialize angular rate states. This configuration reflects realistic onboard sensing and computation, unlike the vast majority of the existing literature that relies on high-precision motion capture systems for state measurement. For comparison purposes, and in view of the time-varying reference profile considered, we additionally implemented the LBC and LMPC controllers. The computational results in Section~\ref{subsec:simulations} indicate that RTMPC and RATMPC are not suitable for real-time implementation, and furthermore, these schemes do not accommodate time-varying references.

\begin{figure}[btp]
\centerline{\includegraphics[width=10em]{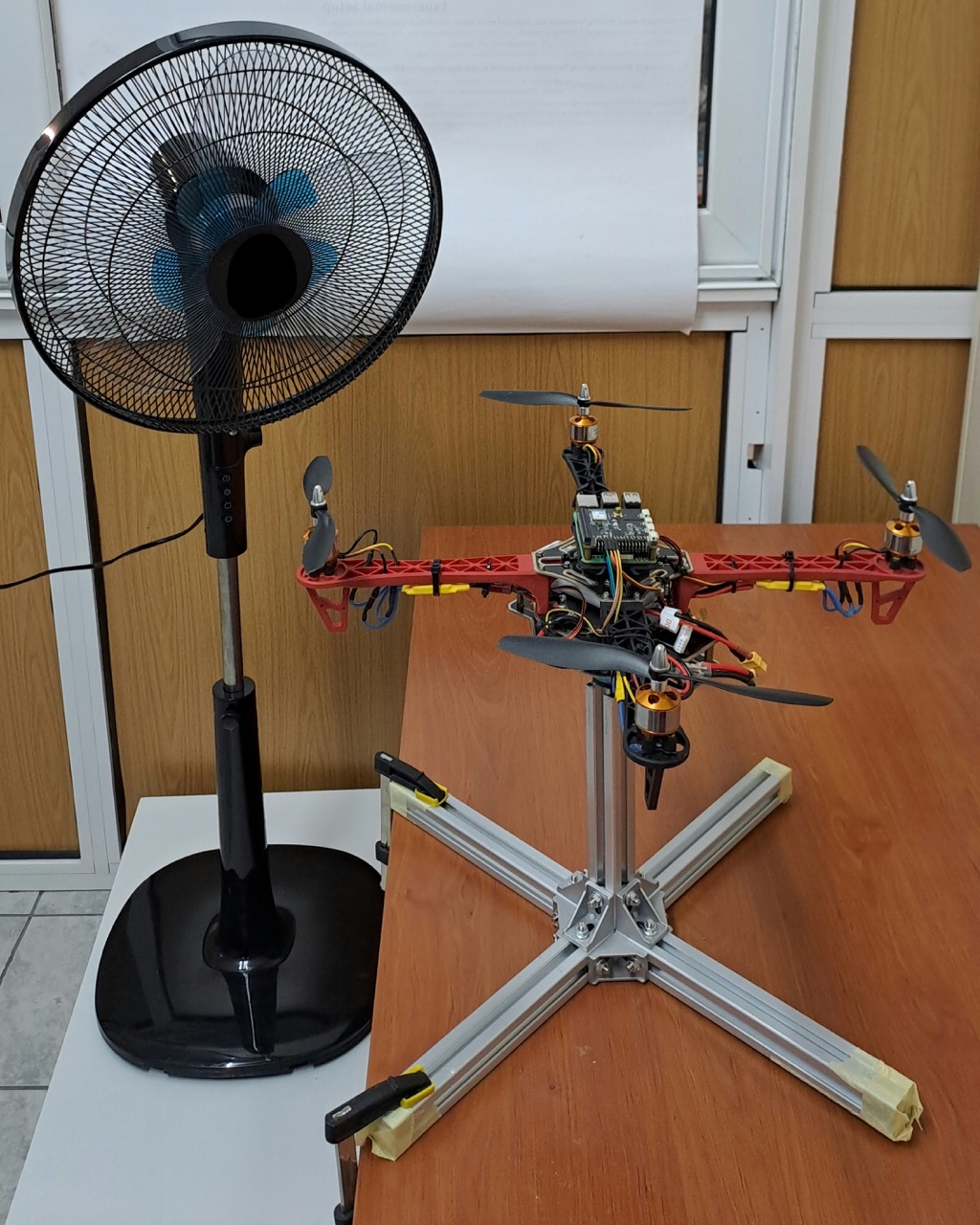}}
\caption{The custom-built quadcopter platform}
\label{fig:quadcopter}
\vspace{-0.25cm}
\end{figure}

The initial angles of the quadcopter are approximately zero and a time-varying reference profile is defined for each attitude channel, consisting of an initial sinusoidal segment followed by two quintic polynomial trajectories. This selection is intended to evaluate the proposed RALMPC framework under both dynamic trajectory tracking and stationary setpoint stabilization conditions; the detailed description of $r(t)$ is given in appendix \ref{app:exp-reference}. For safety reasons, the angular rate constraints are set to $\{|\dot{\phi}|,|\dot{\theta}|,|\dot{\psi}|\}\leq 50^{o/s}$. The sampling time for RALMPC and LMPC is $0.02\text{sec}$ and for LBC $0.01\text{sec}$. The RALMPC horizon is set to $N_p=10$ and the remaining controller parameters are provided in Table \ref{tab:exp-case-I-controller-parameters}, while the parameters for LBC and LMPC were fine-tuned to optimize their closed-loop responses; the parameters related to disturbance estimation were set to $0$, since there is no external disturbance. Note that in this case study, different controller parameters were selected for each axis due to the discrepancies that inherently exist between the rotational motions and that are caused by the hardware parts.

\bgroup
\def\arraystretch{1.3}
\begin{table}[btp]
\centering
\begin{threeparttable}
\caption{Experiment - Case I: RALMPC Parameters}
\label{tab:exp-case-I-controller-parameters}
\setlength{\tabcolsep}{3pt}
\begin{tabular}
{P{25pt}P{12pt}P{12pt}P{15pt}P{70pt}P{10pt}P{10pt}P{10pt}P{15pt}}
\hline \hline
Axis & $Q_1$ & $Q_2$ & $R_1$ & $\gamma_i,i=\{1,2,3\}$ & $k_1$ & $k_2$ & $\phi_p$ & $\sigma_m$ \\ \hline
Roll & $20$ & $2$ & $10^{-3}$ & $\{0.25,0.1,0.1\}$ & $2.5$ & $0.5$ & $1$ & $10^{-3}$ \\
Pitch & $30$ & $1.5$ & $10^{-3}$ & $\{0.25,0.1,0.1\}$ & $2.5$ & $0.5$ & $1$ & $10^{-3}$ \\
Yaw & $75$ & $0.5$ & $10^{-3}$ & $\{0.25,0.2,0.1\}$ & $2.5$ & $0.5$ & $1$ & $10^{-3}$ \\
\hline \hline
\end{tabular}
\end{threeparttable}
\vspace{-0.25cm}
\end{table}

\subsubsection{Results}
\label{subsubsec:exp-results}
Table \ref{tab:exp-case-I-performance metrics} presents the closed-loop evaluation metrics; similarly to the simulation scenario, it is evident that RALMPC consistently outperforms its rivals, as it achieves lower MAE values alongside reduced control effort across all rotational axes. Furthermore, Fig. \ref{fig:exp-case-I-axes} depicts the angular displacements and torque inputs across all axes for the considered controllers. Figs. \ref{fig:exp-case-I-contractive-constraint}-\ref{fig:exp-case-I-solver-info} depict the value of the contractive constraint, the RBFNN weights' $\mathcal{L}_2$-norm, moments of inertia (MoI) and $\hat{\epsilon}$ estimates, and the cost function evolution together with the computational time required to solve the OCP.

From Fig. \ref{fig:exp-case-I-axes}, it is evident that RALMPC outperforms its rivals, as LBC exhibits oscillatory behavior during the initial transient part and larger deviations from the reference during the segments, where the reference changes from sinusoidal to polynomial. Compared to LMPC, it is apparent that RALMPC achieves an overall superior performance, as LMPC exhibits a large constant offset, which is due to its non-adaptive nature. Note that the slight oscillatory behavior observed across all axes for the entire operation exists because of the inherent measurement noise and because the direct on-board measurements are fed to the controllers. The latter is quite important, as it showcases that RALMPC is capable of achieving high accuracy, while operating under realistic conditions. In addition, note that the aforementioned are achieved, while simultaneously satisfying the imposed constraints in terms of states and control inputs. The latter validates that even under the relaxed input constraints imposed through the RPBF \eqref{eq:deadzone_barrier}, RALMPC is capable of guaranteeing safety, while optimizing the closed-loop performance in parallel.

Furthermore, Fig. \ref{fig:exp-case-I-contractive-constraint} demonstrates that the values of the contractive constraint \eqref{eq:ocp-contractive-constraint} remain negative. More specifically, notice that especially during the initial transient and the segments between the sinusoidal and polynomial parts, RALMPC drives the contractive constraint to more negative values indicating a faster convergence rate of the tracking errors to the invariant sets given by \eqref{eq:theorem-1-second-argument}. Regarding the adaptive quantities, from Fig. \ref{fig:exp-case-I-est_quant}, it is evident that they continuously adapt to compensate the uncertainties, ultimately leading to improved closed-loop performance compared to LMPC. Fig.~\ref{fig:exp-case-I-solver-info} depicts the cost function, which is driven to a neighborhood of the origin, and the time required to solve the OCP, respectively; the latter is at the order of a few milliseconds with a maximum value in the initial steps around 5ms, thus validating the computational efficiency of RALMPC under realistic conditions. An indicative video of the experiments can be found online\footnote{An indicative video of the experiments is available at \url{https://youtu.be/AkTAkm8D4iw}\label{videofoot}.}.

\bgroup
\def\arraystretch{1.4}
\begin{table}[btp]
\vspace{-0.5cm}
\centering
\begin{threeparttable}
\caption{Experiment - Case I: Performance Metrics}
\label{tab:exp-case-I-performance metrics}
\setlength{\tabcolsep}{4pt}
\begin{tabular}{P{40pt}P{25pt}P{25pt}P{25pt}P{25pt}P{25pt}P{25pt}}
\hline \hline
\textbf{Controller} & $\text{MAE}_{\phi}$ & $\text{MAE}_{\theta}$ & $\text{MAE}_{\psi}$ & $||\tau_\phi||_{2}$ & $||\tau_\theta||_2$ & $||\tau_\psi||_2$ \\[2pt] \hline \hline
\textbf{RALMPC} & $0.7763$ & $0.4801$ & $0.6677$ & $7.0767$ & $5.6763$ & $12.0963$ \\ 
\textbf{LBC} & $1.4847$ & $1.2421$ & $1.2096$ & $7.1341$ & $6.8373$ & $14.5634$ \\
\textbf{LMPC} & $5.2655$ & $3.4232$ & $2.9189$ & $8.8656$ & $7.3153$ & $12.2603$ \\
\hline \hline
\end{tabular}
\end{threeparttable}
\end{table}

\begin{figure}[t]
\centerline{\includegraphics[width=\columnwidth,height=5.75cm]{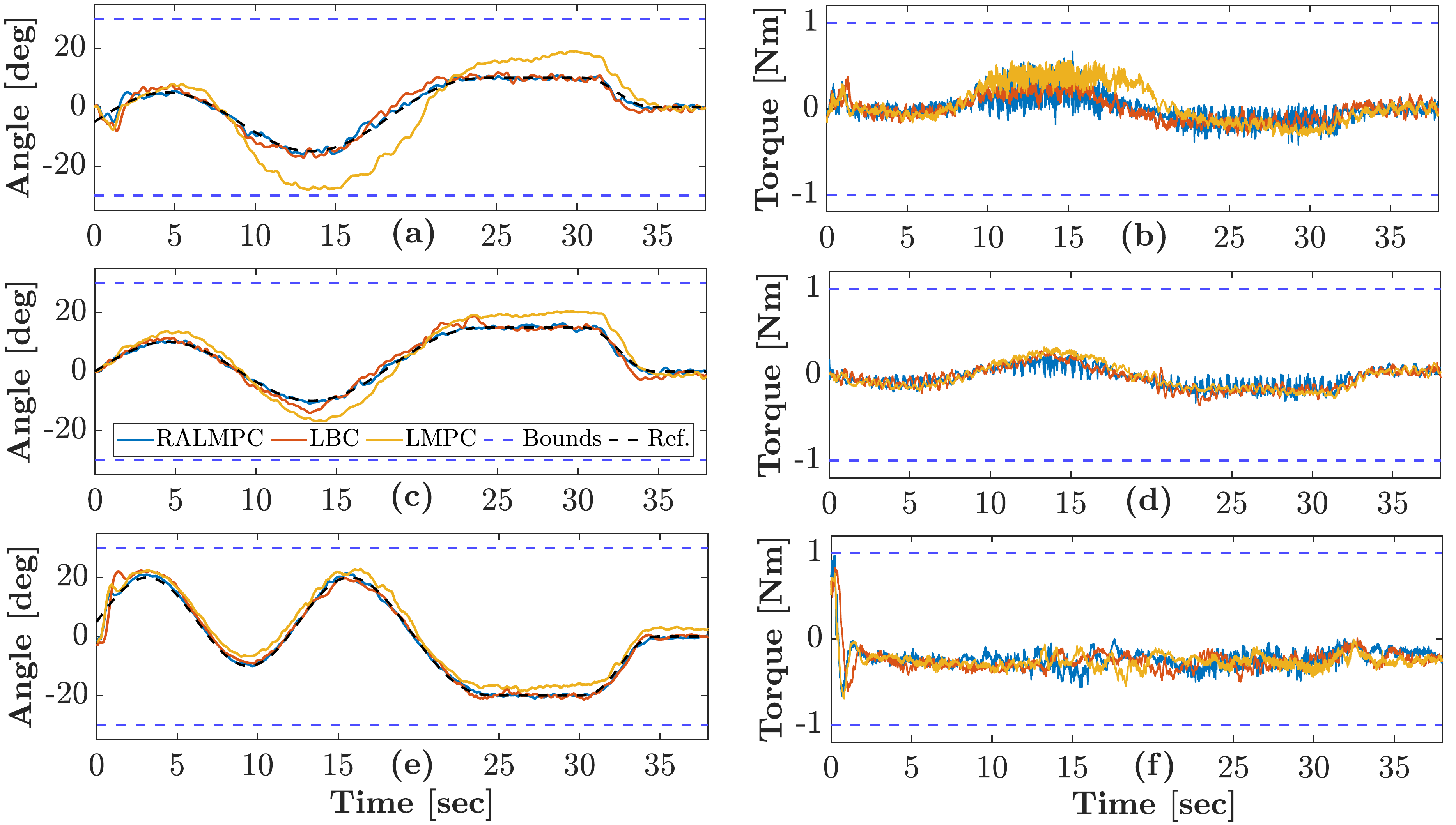}}
\caption{Experiment - Case I: Roll axis: (a) Angle - (b) Torque -- Pitch axis: (c) Angle - (d) Torque -- Yaw axis: (e) Angle - (f) Torque}
\label{fig:exp-case-I-axes}
\vspace{-0.5cm}
\end{figure}

\begin{figure}[t]
\centerline{\includegraphics[width=\columnwidth]{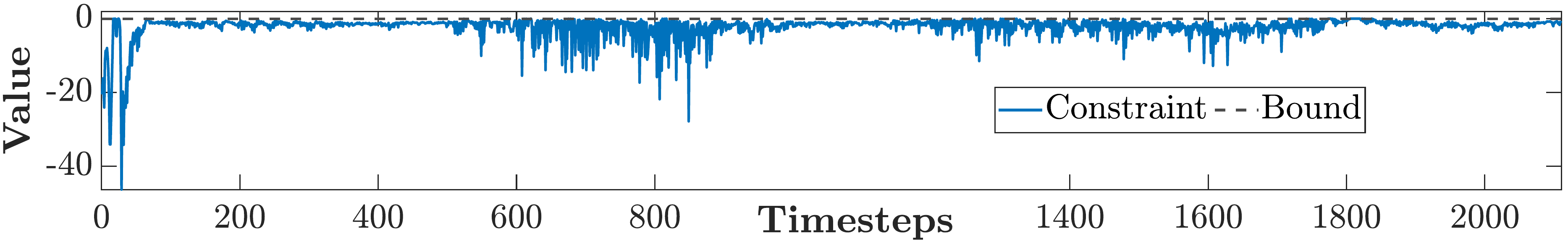}}
\caption{Experiment - Case I: Contractive constraint value}
\label{fig:exp-case-I-contractive-constraint}
\vspace{-0.25cm}
\end{figure}

\begin{figure}[t]
\centerline{\includegraphics[width=\columnwidth]{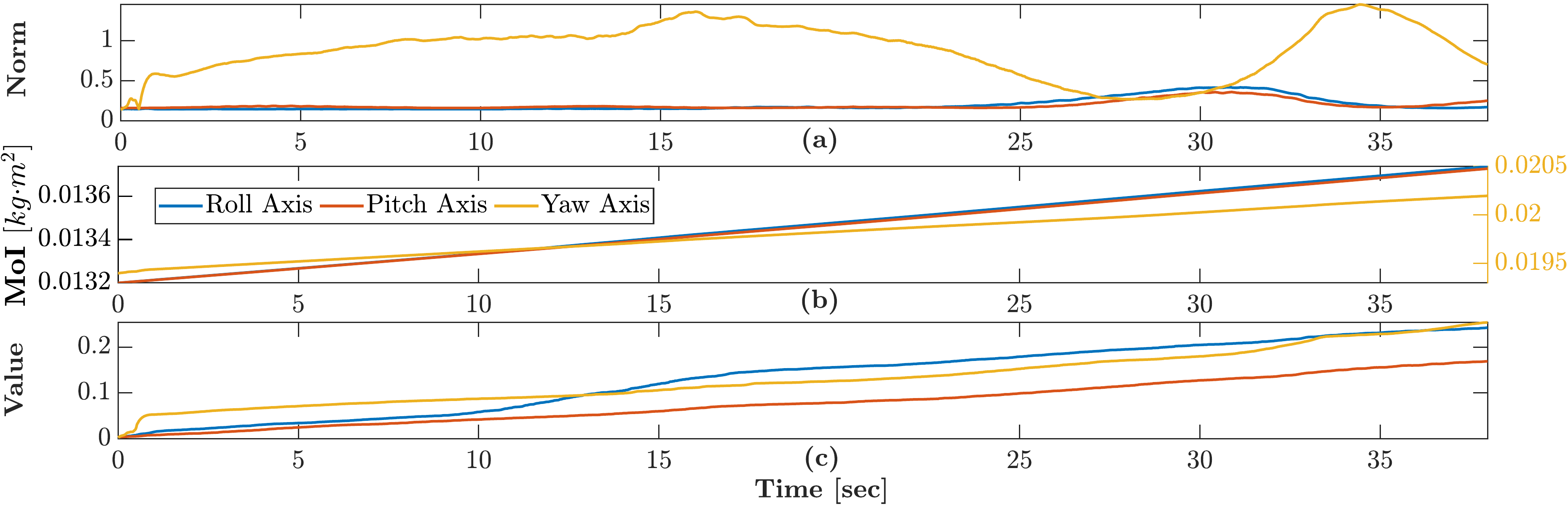}}
\caption{Experiment - Case I: (a) RBFNN weights norm - (b) MoI estimates - (c) $\hat{\epsilon}$ estimates}
\label{fig:exp-case-I-est_quant}
\vspace{-0.35cm}
\end{figure}

\begin{figure}[t]
\centerline{\includegraphics[width=\columnwidth]{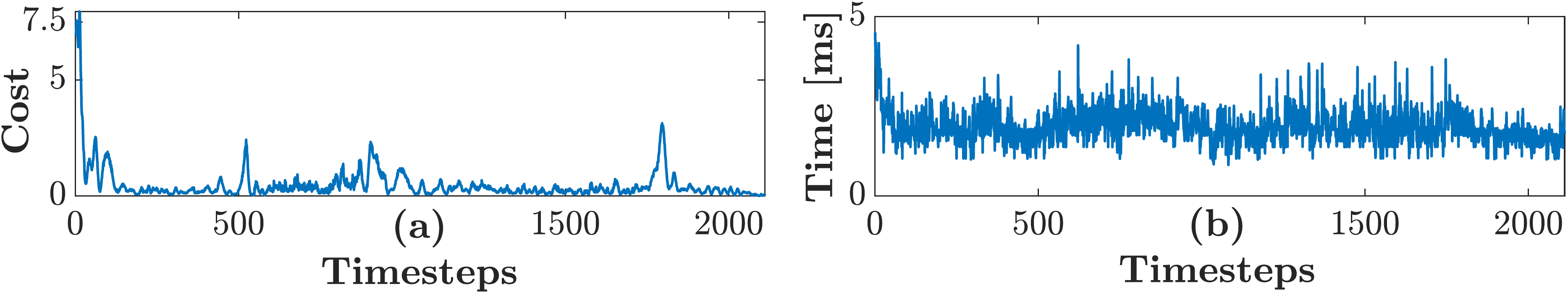}}
\caption{Experiment - Case I: (a) Cost function - (b) OCP solution time}
\label{fig:exp-case-I-solver-info}
\vspace{-0.25cm}
\end{figure}

\vspace{-0.3cm}
\subsection{Experiment - Case II}
\label{subsec:exp-case-II-dist}
\subsubsection{Setup Description}
\label{subsubsec:exp-case-II-setup}
In this case study, we deploy RALMPC to track the same trajectory as in the previous case study, but under the influence of external disturbances injected by a fan, as shown in Fig. \ref{fig:quadcopter}. The controller parameters are the same as in Table \ref{tab:exp-case-I-controller-parameters} and the adaptation rate of the disturbance estimation is $\gamma_4=0.05$. To comprehensively evaluate the disturbance rejection capabilities, we present two distinct scenarios: i) sudden step disturbance: the fan operating at maximum capacity is abruptly directed toward the quadcopter, simulating a sudden, high-magnitude wind gust, and ii) gradual ramp disturbance: the fan is continuously directed at the quadcopter while its aerodynamic output is progressively increased from zero to maximum capacity, simulating a slowly varying disturbance. The former scenario corresponds to a pure piecewise-constant disturbance aligned with assumption 3, while the latter scenario is considered to satisfy $\dot{d}(t)\approx 0$.

\subsubsection{Results}
\label{subsubsec:exp-case-II-results}
Figs. \ref{fig:exp-case-II-axes} and \ref{fig:exp-case-II-ramp-axes} depict the angles and torque inputs for the two disturbance scenarios, respectively. From Fig. \ref{fig:exp-case-II-axes}, it is evident that the imposed step disturbance, which is injected around $t=6\text{sec}$ leads to a slight and more pronounced oscillatory behavior in the pitch axis, while the influence on the yaw channel results in an offset from the reference. However, across all axes the deviations of the angles from the reference are rapidly compensated by the controller, which is also evident from the torque profiles that exhibit an apparent modification of their trend during and after disturbance injection. Consequently, the closed-loop responses are driven back to a small neighborhood of the reference and remain there for the rest of the operation. Fig.~\ref{fig:exp-case-II-ramp-axes} shows that the proposed RALMPC effectively compensates for the time-varying disturbance affecting the quadcopter from the onset of the operation, as it consistently maintains the angles within a small neighborhood of the reference without pronounced deviations throughout the experiment. Finally, note that under all circumstances the state and control constraints are satisfied.

\begin{figure}[btp]
\centerline{\includegraphics[width=\columnwidth]{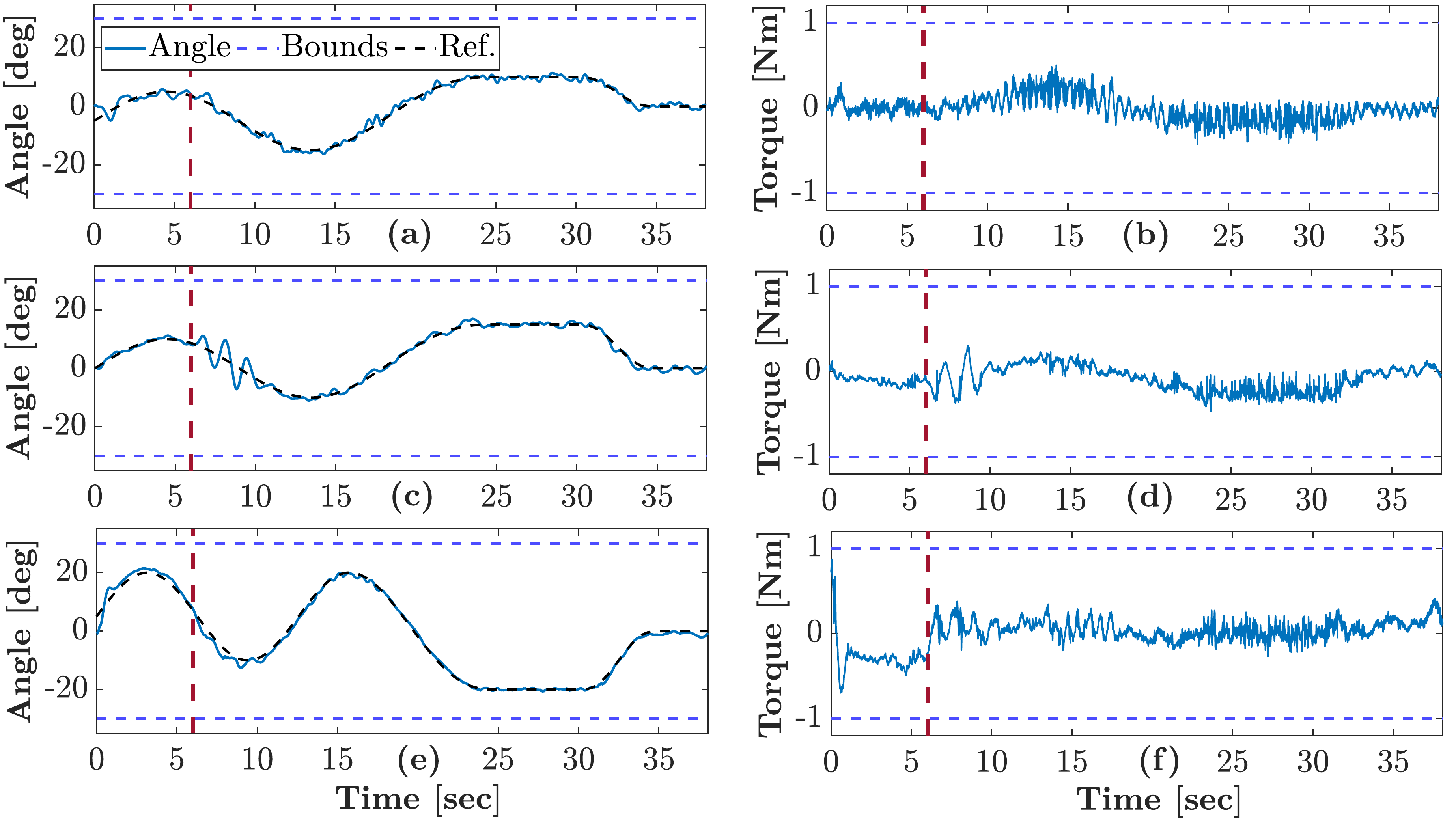}}
\caption{Experiment - Case II: Step disturbance: Roll axis: (a) Angle, (b) Torque -- Pitch axis: (c) Angle, (d) Torque -- Yaw axis: (e) Angle, (f) Torque -- The red dashed line indicates the disturbance injection time.}
\label{fig:exp-case-II-axes}
\vspace{0.5cm}

\centerline{\includegraphics[width=\columnwidth]{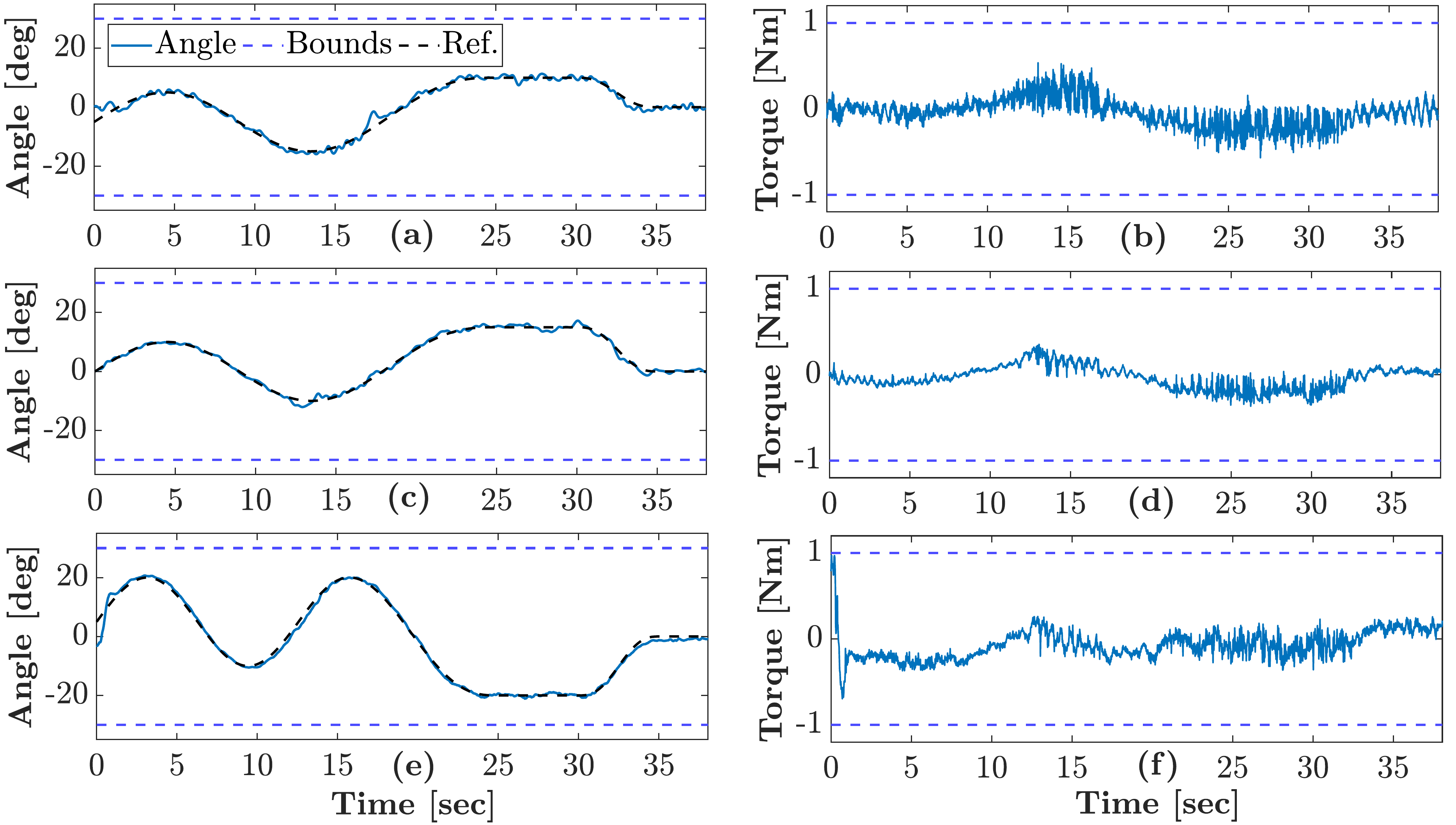}}
\caption{Experiment- Case II: Gradual Ramp disturbance: Roll axis: (a) Angle, (b) Torque -- Pitch axis: (c) Angle, (d) Torque -- Yaw axis: (e) Angle, (f) Torque}
\label{fig:exp-case-II-ramp-axes}
\vspace{-0.5cm}
\end{figure}

\vspace{-0.2cm}
\section{Conclusions}
\label{sec:conclusions}
This paper proposes a robust adaptive Lyapunov-based MPC (RALMPC) framework for time-varying output tracking of uncertain nonlinear constrained systems. An iBLF-based backstepping controller with RBFNN approximation is developed to compensate for parametric and non-parametric uncertainties and external disturbances, while guaranteeing constraint satisfaction, uniform boundedness, and convergence of the tracking errors to compact sets. The derived Lyapunov-based laws are used to construct a contractive constraint for the RALMPC OCP, ensuring recursive feasibility and inheritance of the closed-loop properties without robust invariant tubes or stabilizing terminal ingredients. Online adaptation is incorporated in the prediction model, and an RPBF was used to account for actuation limits without compromising feasibility. Hardware-in-the-loop simulations and experimental results on quadcopter attitude control demonstrate improved computational efficiency, closed-loop performance, and real-time applicability. Future work will extend the framework to broader system classes.

\vspace{-0.2cm}
{\appendices
\section{Quadcopter Attitude Nonlinearities}
\label{app:quad-unknown-dynamics}
This appendix derives the nonlinear roll-axis term $F_\phi(x)$ in \eqref{eq:attitude-model}. Although treated as unknown dynamics in the simulation study, these terms capture physical effects that perturb real quadcopters. The pitch and yaw derivations are omitted, as they follow analogously. The values of the various coefficients used for the simulations are provided in Table \ref{tab:unknown-dynamics-coefficients}.

\begin{enumerate}[leftmargin=*,label=(\arabic*)]
    \item Direct roll-rate damping $F_{damp}(x)=-d_1\cdot p-d_2\cdot|p|\cdot p \label{eq:roll-rate-damping}$, where $d_1,d_2>0$. This term captures viscous- and drag-like damping effects.

    \item Cross-axis rate-dependent damping $F_{cr}(x)=-d_3\cdot|q|\cdot p-d_4\cdot|r|\cdot p \label{eq:roll-cross-damping}$, where $d_3,d_4>0$. This terms represent dissipation depending on the other axes.

    \item Residual bilinear in rates cross-axis coupling $F_{cpl}(x)=c_1\cdot p\cdot q + c_2\cdot p\cdot r + c_3\cdot q\cdot r $, where $c_i,i=1,2,3$ are constant terms. Beyond the nominal $\frac{I_\theta - I_\psi}{I_\phi}qr$ term, additional couplings arise due to parameter mismatch, flexibility, rotor-wake interactions and neglected terms.

    \item Quadratic cross-axis coupling via absolute values $F_{abs,cpl}(x) = c_4\cdot |q|\cdot p + c_5\cdot |r|\cdot p + c_6\cdot |q|\cdot r$, where $c_i,i=4,5,6$ are constants. These terms emulate drag-like cross couplings that scale with the magnitude of other-axis rates. The first two terms resemble coupling-induced damping/gain modulation, while the last term magnitude-based cross coupling.

    \item Attitude-dependent parasitic moments $F_{att}(x)=a_1\cdot\sin(\phi) + a_2\cdot\sin(\theta) + a_3\cdot \sin(\psi)$, where $a_i>0,i=1,2,3$. These terms capture unmodeled attitude-related moments (mass imbalance, aero asymmetry).

    \item Attitude-rate mixed parasitics $F_{att,rate}(x) =  a_4\cdot\sin(\phi)\cdot p + a_5\cdot\sin(\theta)\cdot q + a_6\cdot\sin(\psi)\cdot r$, where $a_i>0,i=\{4,5,6\}$. These terms model angle-modulated damping and orientation-dependent aerodynamic loading.

    \item Gyroscopic/rotor-induced effects $F_{gyro}(x) = g_1\cdot q + g_2\cdot r + g_3\cdot q\cdot r + g_4\cdot p\cdot q\cdot r$, where $g_i,i=\{1,\dots,6\}$ are constants. Even though true gyroscopic torques depend on rotor speeds, these state-dependent terms capture rate-coupled effects.

    \item Higher-order polynomial nonlinearities $F_{poly}(x) = k_1\cdot p^3 + k_2\cdot p\cdot q^2 + k_3\cdot p\cdot r^2 + k_4\cdot q^3 + k_5\cdot r^3$ where $k_i{>}0,i=1,\dots,5$. These terms emulate higher-order unmodeled effects (nonlinear drag, structural nonlinearities).
\end{enumerate}

\bgroup
\def\arraystretch{1.4}
\begin{table}[btp]
\centering
\begin{threeparttable}
\caption{Quadcopter unknown dynamics: Coefficient values}
\label{tab:unknown-dynamics-coefficients}
\setlength{\tabcolsep}{4pt}
\begin{tabular}{P{15pt}P{20pt}P{15pt}P{15pt}P{15pt}P{20pt}P{15pt}P{20pt}P{20pt}P{17pt}}
\hline \hline
$d_1$ & $d_2$ & $d_3$ & $d_4$ & $c_1$ & $c_2$ & $c_3$ & $c_4$ & $c_5$ & $c_6$ \\[2pt]
$1.5$ & $0.075$ & $0.3$ & $0.3$ & $0.3$ & ${-}0.25$ & $0.15$ & $0.1$ & ${-}0.1$ & $0.08$ \\ \hline
$a_1$ & $a_2$ & $a_3$ & $a_4$ & $a_5$ & $a_6$ & $g_1$ & $g_2$ & $g_3$ & $g_4$ \\
$0.8$ & $0.5$ & $0.2$ & $0.1$ & $0.08$ & $0.06$ & $0.15$ & ${-}0.12$ & $0.05$ & $0.005$ \\ \hline
\multicolumn{2}{c}{$k_1$} & \multicolumn{2}{c}{$k_2$} & \multicolumn{2}{c}{$k_3$} & \multicolumn{2}{c}{$k_4$} & \multicolumn{2}{c}{$k_5$} \\
\multicolumn{2}{c}{$0.003$} & \multicolumn{2}{c}{$0.0015$} & \multicolumn{2}{c}{$0.0015$} & \multicolumn{2}{c}{$0.001$} & \multicolumn{2}{c}{$0.001$} \\
\hline \hline
\end{tabular}
\end{threeparttable}
\vspace{-0.5cm}
\end{table}

\section{Reference Trajectory of Experiment}
\label{app:exp-reference}
The full reference trajectory for each axis can be expressed compactly in piecewise form for $i=\{\phi,\theta,\psi\}$ as\begin{align*}
r_i(t) =
\begin{cases}
b_i + A_i \sin(\omega_i t), 
& t \in [0 \ t_1) \\[2pt]
\displaystyle
\sum_{k=0}^{5} c_{k,i}^{(11)}
\left( \frac{t - t_1}{T_{11}} \right)^k,
& t \in [t_1 \ t_2) \\[4pt]
y_{2,i},
& t \in [t_2 \ t_3) \\[2pt]
\displaystyle
\sum_{k=0}^{5} c_{k,i}^{(12)}
\left( \frac{t - t_3}{T_{12}} \right)^k,
& t \in [t_3 \ t_4) \\[4pt]
0,
& t \ge t_4
\end{cases}
\end{align*}
where $c_{k,i}^{(11)}$ are the coefficients of the first quintic polynomial that transitions the reference from the sinusoidal segment to the intermediate setpoint $y_{2,i}$, $c_{k,i}^{(12)}$ are the coefficients of the second quintic polynomial that transitions the reference from $y_{2,i}$ to the origin, and  $T_{11},T_{12}>0$ are the durations of the transition intervals for the first and second polynomial, respectively. The coefficients of each polynomial are computed based on the continuity requirements that are defined as follows by considering $r_i(t)=p_1(s),t\in[t_1,t_2)$ with $s=\frac{t-t_1}{T_{11}}$ and $r_i(t)=p_2(s),t\in[t_3,t_4)$ with $s=\frac{t-t_3}{T_{12}}$:\begin{align*}
    &p_1(0) = r_{i,1} & \frac{\dot{p}_1(0)}{T_{11}} &= v_{i,1} & \frac{\ddot{p}_1(0)}{T_{11}^2} &= a_{i,1} \\
    &p_1(1)=y_{2,i} & \frac{\dot{p}_1(1)}{T_{11}}&=0 & \frac{\ddot{p}_1(1)}{T_{11}^2} &=0
\end{align*}where $r_{i,1},v_{i,1},a_{i,1}$ are the position, velocity and acceleration values of the sinusoidal segment at $t{=}t_1$; the coefficients for the second polynomial are calculated similarly. We considered $b_i=[10 \ 0 \ 5]^T\text{deg},A_i=[10 \ 10 \ 15]^T\text{deg},\omega_i=[0.35 \ 0.35 \ 0.5]^T\text{rad/sec},t_1=15\text{sec},T_{11}=10\text{sec},\Delta t_{23}=t_3-t_2=5\text{sec},y_{2,1}=[10 \ 15 \ {-}20]^T\text{deg},T_{12}=5\text{sec}$.
}


 
%
\balance

\end{document}